\documentclass{aastex6}
\usepackage{amsmath}
\usepackage{amssymb}
\usepackage[normalem]{ulem}
\usepackage{appendix}
\usepackage{graphicx}
\usepackage{natbib}

\def\numax{$\nu_{\rm max}$}

\def\dnu{$\Delta\nu$}
\def\teff{$T_{\rm eff}$}

\def\be{\begin{equation}}
\def\ee{\end{equation}}
\def\bea{\begin{eqnarray}}
\def\eea{\end{eqnarray}}
\def\feh{$\mathrm{[Fe/H]}$}
\def\eps{$\epsilon$}
\def\msun{M$_\odot$}
\def\rsun{R$_\odot$}

\newcommand{\overbar}[1]{\mkern 1.5mu\overline{\mkern-1.5mu#1\mkern-1.5mu}\mkern 1.5mu}

\begin{document}

\shorttitle{Robustness of asteroseismic results}
\shortauthors{Basu \& Kinnane}


\title{The robustness of asteroseismic estimates of global stellar parameters\\
to surface term corrections}

\author{Sarbani Basu, and Archer Kinnane}

\affil{Department of Astronomy, Yale University, New Haven, CT, 06520, USA}

\email{sarbani.basu@yale.edu}

\begin{abstract}
Oscillation frequencies of even the best stellar models differ from those of the stars they represent,
and the difference is predominantly a function of frequency.
This difference is caused by limitations of modeling the
near-surface layers of a star. This frequency-dependent frequency error, usually referred
to as the ``surface term'' can result in erroneous interpretation of the oscillation
frequencies unless treated properly.  Several techniques have been developed
to minimize the effect of the surface term; it is either subtracted out, or frequency
combinations insensitive to the surface terms are used, or the asteroseismic phase \eps\
is used to determine a match between models and stars.  In this paper we show that 
no matter
what technique is used to account for the
surface term, as long as the physics of the models is the same,
 the global parameters of a star --- mass, radius and, age --- 
obtained from frequency analyses are robust. This implies that even though we cannot
model the internal structure of stars perfectly, we can have confidence in all 
results that use stellar global properties obtained through the analysis of stellar
oscillation frequencies.
\end{abstract}

\keywords{stars: oscillations --- stars: fundamental parameters
--- methods: statistical}

\section{Introduction}
\label{sec:intro}

The availability of asteroseismic data has changed the way we determine the properties of
stars. These data allow us to determine the radii, masses and ages of stars to
an unprecedented degree of precision \citep[see e.g.,][]{chaplin2014, serenelli2017,
victor2015, victor2017, pin2014, pin2018}. 
While reasonable results are obtained when only 
average seismic data --- the large separation \dnu\ and the frequency of maximum
power \numax\ --- are available  the best results are obtained
when data on individual frequencies are available.

Like most of astrophysics, analysis of asteroseismic data to determine stellar properties
is done by comparing the properties of the models to the observed properties
of the star under consideration. In this case we compare the frequencies
of the models that also have  the same effective
temperature and metallicity as the star.  There is however, a complication in this, and
that is the so-called ``surface term''.
The surface term is a frequency-dependent frequency difference between stellar models and stars that
arises from our inability to model the near-surface layers of a star properly \citep{brown1984, jcdetal1988}.
The surface term can also be seen between models constructed with different surface
physics \citep[e.g.][and references therein]{basu2016}.
For modes of
degree $l \lesssim 200$, the surface term is independent of the degree of the mode \citep{basu1996}.

Asteroseismic analyses have to take the surface term into account somehow. Analyses done
only with average asteroseismic properties \dnu\ and \numax\ generally ignore the surface term
\citep[e.g., the analyses of][]{chaplin2014, pin2014}, though recent analyses assume
that the surface term contribution to \dnu\ scales as the surface term for the 
Sun \citep{serenelli2017}. For analyses where individual modes were used \citep[e.g.][]{metcalfe2010, metcalfe2012,
victor2015, victor2017}, a variety of different ways have been used to remove the
effects of the surface term.

The aim of this paper is to examine whether the global properties of a star --- mass, radius, and age ---
are affected by how the surface term is handled.  For this we 
analyze solar frequencies degraded to
stellar levels, the frequencies of 16Cyg~A and B, as well as those of
 KIC~6106415 (HD~177153), KIC~6225718 (HD~187637), and KIC~8006161 (HIP~91949) using a variety
of different methods. We should note that our investigation is limited to studying the
effects of the surface term on the result.  It is known that the results, in particular, that for
age, can change depending on the physics of the models \citep[see e.g.,][]{lebreton2014b}; mass
and radius results are much more robust to changes in physics \citep{ning, victor2015, victor2017}.
It is difficult to gauge the effect of the surface term on results from previous analyses
such as those of \citet{metcalfe2010}, \citet{victor2015}, \citet{victor2017}
etc., since the use of different surface terms was usually combined with the use of different
physics inputs in the models.

The rest of the paper is organized as follows:
In Section~\ref{sec:surf} we discuss the surface term in more detail and discuss some of the
common ways of accounting for it while making asteroseismic estimates of 
stellar properties. In Section~\ref{sec:method} we describe the analysis techniques in detail and list the different
ways we determined the global stellar parameters. We describe the stars and their data in Section~\ref{sec:stars}. 
We describe how we constructed models and the parameter ranges used in Section~\ref{sec:models}.
We present our results in Section~\ref{sec:res}. Our conclusions are stated in Section~\ref{sec:conclu}.

\section{The surface term: sources and mitigation}.
\label{sec:surf}

\begin{figure}
\includegraphics[width=2.75 true in]{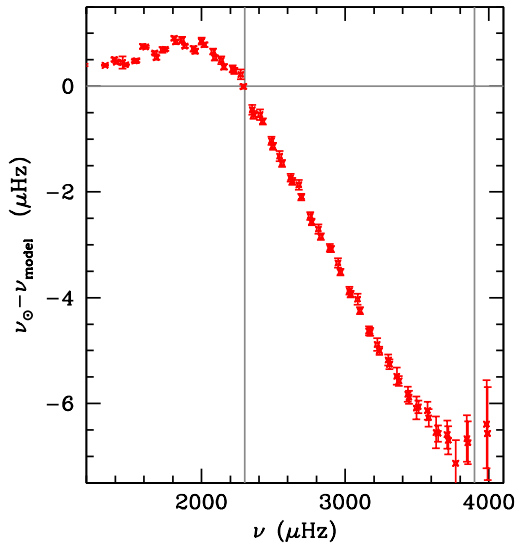}
\caption{The frequency differences  between the Sun and standard solar model BS05(OP) of
\citet{jnb2005}. The solar frequencies are those from the Birmingham Solar Oscillation
Network (BiSON) published by \citet{wjc2007}. The vertical gray lines mark the frequency
range of modes that could be obtained for the Sun by a mission like {\it Kepler}.}
\label{fig:ssmsurf}
\end{figure}

The surface term plagues helioseismic analyses too. We show the
surface term between a standard solar model (SSM) and the Sun in Fig.~\ref{fig:ssmsurf}.
The surface term arises from a number of factors, but chief among them is the
treatment of convection in models. Usual one-dimensional stellar models treat
convection in a very simplified manner using approximations of like the mixing-length theory 
(MLT; \citealt{bohmvitense1958}) or variants thereof (e.g., \citealt{cm1991}, \citealt{arnett2010}). While
these approximations work well in the deeper parts of the convection zone where 
convection is efficient, they do not model the near-surface superadiabatic layer
very well. Also missing in these models are dynamical effects of convection, in 
particular pressure support provided by turbulence --- turbulent pressure can
be as high as  15\% of the gas pressure in the superadiabatic regions (\citealt{stein1998}, 
\citealt{nord1997}, \citealt{tanneretal2013}). Other limitations of near-surface modelling 
include the use of very simple models of stellar atmospheres, as well as 
uncertainties in microphysics inputs such as low-temperature opacities. Another
contribution to the surface term comes from the fact that the adiabatic approximation for
treating stellar oscillation frequencies breaks down near the surface; frequencies
are usually calculated assume that the linear adiabatic wave equations apply, while
there are codes (e.g., Gyre; \citealt{gyre}) that can determine non-adiabatic
frequencies, they do not usually include the effect of convective heating and cooling
on oscillation modes. All factors mentioned above affect very shallow
layers of stars, and it can be shown from the theory of oscillations that perturbations
to near-surface layers cause a frequency-dependent frequency change that is a 
smooth function of frequency that can be represented
as a low-degree polynomial (see Fig. 3.17 of \citealt{mybook}).

There have been several concerted efforts to improve { the near-surface structure of models} in order to 
decrease the surface term. \citet{rosenthal1995} showed that solar models
constructed with  non-local mixing length formulation of 
\citet{dog1977} as formulated by \citet{balmforth}  reduces the surface term with respect to the Sun. 
\citet{demarque1997} showed that parametrizing the structure of the super-adiabatic layer 
in simulations and applying that to solar models improves frequencies.  
Other efforts
include patching the results of realistic
convection simulations on stellar models \citep[e.g.][]{rosenthal, piau2014, joergensen2018}.
\citet{houdek2017} did a detailed study of the physics ingredients that contribute to the
surface term, and they confirmed that there are multiple sources, including effects from
convection dynamics; they show that the surface term can indeed be made negligibly small.
There work, however, was done with solar envelope models and therefore, do
not easily translate to models along an evolutionary sequence.
As a step in this direction, \citep{mosumgaard2018} have been examining models constructed
with near-surface properties obtained by
interpolating withing grids of convection simulations. However, these models still show 
a surface term, probably because \citet{mosumgaard2018} neglected turbulent 
pressure. { However, it is not enough to include turbulent pressure in the models;
\citet{sonoi} showed that to achieve the best results, perturbations to turbulent pressure 
must be taken into account properly when frequencies of such models are calculated.}
Thus even as there are ongoing efforts to improve the near-surface properties
of stellar models, the vast majority of stellar models used have a substantial surface term 
that needs to be accounted for properly.

For helioseismic studies, where many hundreds of modes are available, the problem of
the surface term is handled by relying  on  frequency inversions
rather than frequency comparisons (see \citealt{basu2016} and references therein).
The inversion process explicitly accounts for the surface term \citep[e.g.,][]{dziem1990, 
dziem1991, dappen1991, ab94}
and thus does not depend on correcting the frequencies.
 However, the 
dearth of asteroseismic data (tens of frequencies rather than hundreds) along with
a lack of independent constraints on mass, radius and age, make inversions impractical 
for most stars. As a result, other strategies are required; these involve either
subtracting out the surface term, making it less effective, or using a different observable
related to the frequencies.

\begin{figure*}
\centerline{
\includegraphics[width=3.00 true in]{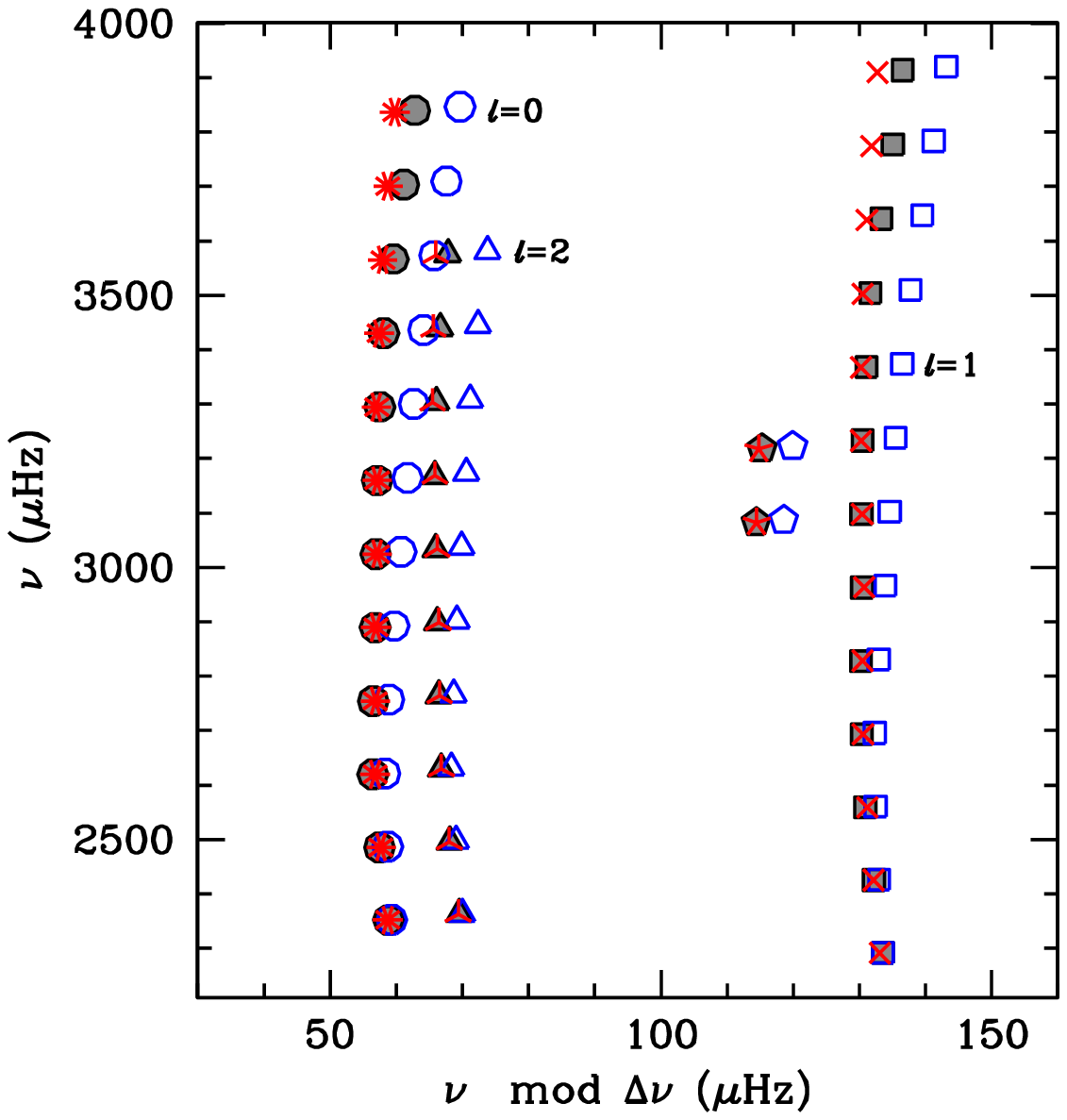}
\includegraphics[width=3.00 true in]{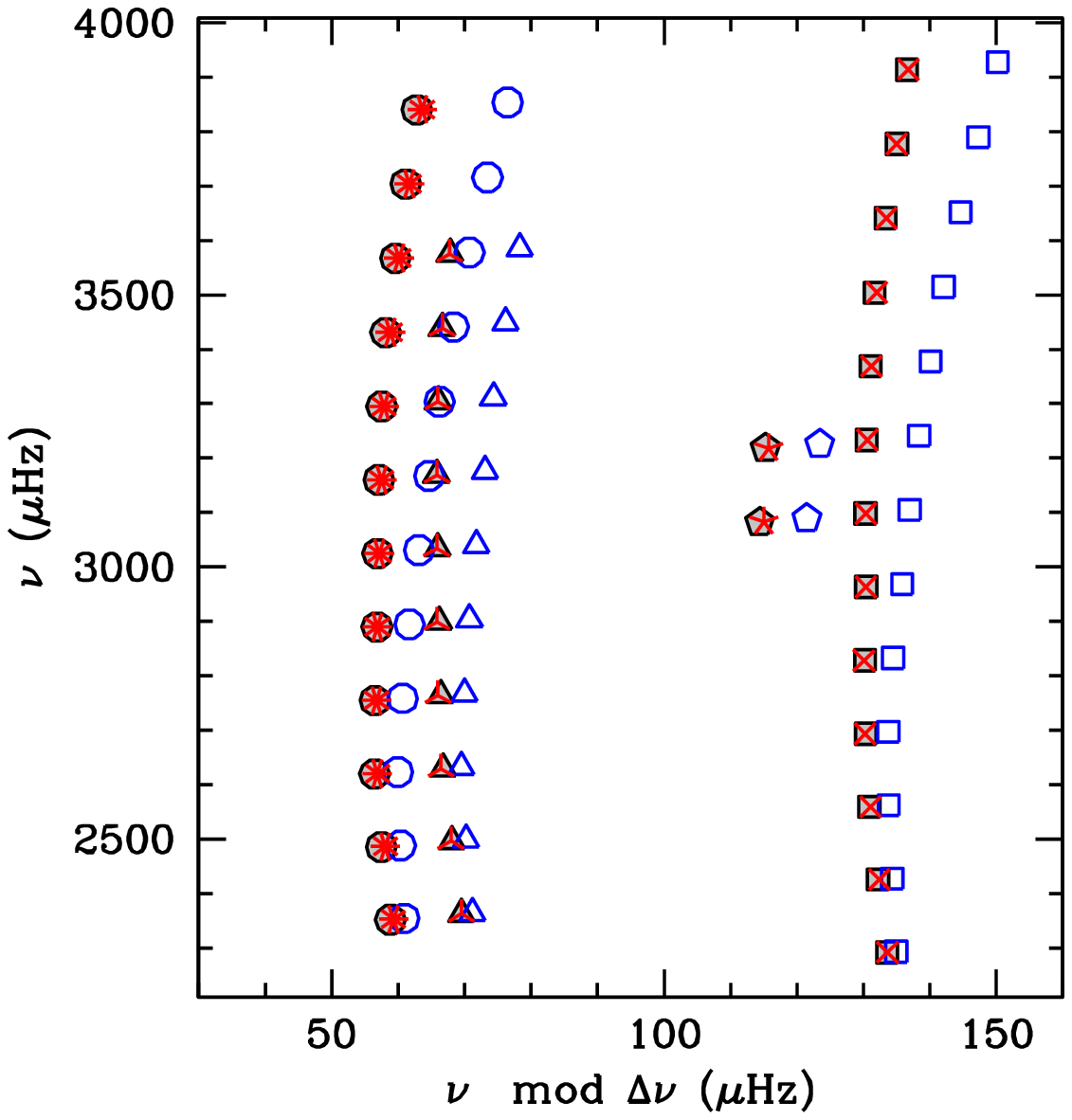}
}
\caption{Echelle diagram showing solar frequencies (black/gray) as well as the uncorrected (blue)
and corrected (red) frequencies of the SSM  BS05(OP) of \citet{jnb2005} (left), and the
best-fit model obtained using the correction in Eq.~\ref{eq:bl} (right). The solar frequencies used
are BiSON data degraded to stellar levels by \citet{lund2017} and used by \citet{victor2017}.
}
\label{fig:sunech}
\end{figure*}

The most common way of removing the effect of the surface term is to assume
that it has a known functional form and subtract it our from the frequency
differences between a star and its model. 
Most early analyses used the surface term correction
proposed by \citet{hans}. That form assumes that the frequency difference caused
by near-surface effects can be expressed as a power-law in frequency, with the exponent
determined from a solar surface term. While this form is easy to use, it depends
on the surface term between the Sun and a solar model, and that depends on the 
physics of the solar model. Additionally, it can be shown (\citealt{joey}) that
this form does not work very well for frequencies far away from \numax, nor does it work
{ too well} for 
models of stars more evolved that the Sun. As a result, it is increasingly common to
use the 
form proposed by \citet{ball2014} based on an
idea of \citet{dog1990}. They proposed that the surface term could be fit to the
form
\be
\delta\nu_{nl} =\nu_{nl}^{\rm obs}-\nu_{nl}^{\rm model}=\frac{1}{I_{nl}}\left[ a\left(\frac{\nu_{nl}}{\nu_{\rm ac}}\right)^{-1}
+b\left(\frac{\nu}{\nu_{\rm ac}}\right)^3\right],
\label{eq:bl}
\ee
where $\delta\nu_{nl}$ is the difference in frequency for a mode of degree $l$ order $n$
between a star and its model, $\nu_{nl}$ is the frequency and  $I_{nl}$ is
the inertia of the mode, and $\nu_{\rm ac}$ is the acoustic cut-off frequency.
The coefficients $a$ and $b$ can be determined through a generalized linear least-squares fit
\citep[see][]{rachel2017}.  The biggest advantage of this form is that one does
not have to rely on a solar model to fix the parameters.  The explicit 
dependence on mode inertia also implies that no special treatment is needed to 
to use the correction on non-radial modes.
This correction works quite well in general, and
we show this in the left column of Fig.~\ref{fig:sunech} where we have corrected the frequencies of 
the standard solar model shown in Fig.~\ref{fig:ssmsurf}; for this purpose we used
 set of solar frequencies degraded to match what is
obtained for other stars with {\it Kepler}. 
Both corrected and uncorrected frequencies are shown for
comparison. 

\begin{figure}
\includegraphics[width=2.75 true in]{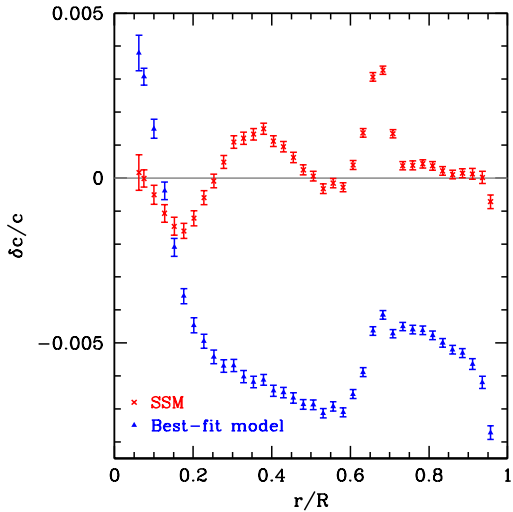}
\caption{The sound-speed difference between the Sun and the SSM BS05(OP) (red) and
the best-fit model (blue). The solar results are from \citet{basu2009}. The
errors bars show $3\sigma$ uncertainty.
}
\label{fig:csqdif}
\end{figure}

The problem with Eq.~\ref{eq:bl} is that it can over-correct the frequencies and remove
differences that are caused by differences in the structure of deeper layers. We show this in the
right column of Fig.~\ref{fig:sunech}, where the corrected and uncorrected frequencies
of a model with $M/M_\odot=1.0142$, $R/R_\odot=1.01$, $T_{\rm eff}=5779$ K and
and age of 4.87~Gyr are plotted; we refer to this model as the ``best-fit''model. 
The corrected frequencies appear to fit the solar data better than those of the 
standard solar model, however the internal structure of best-fit model has a much 
larger mismatch with the Sun than the SSM. This is shown in Fig.~\ref{fig:csqdif}.
Thus clearly, the surface-term correction is giving us misleading results.

As mentioned earlier, it is not usually possible to invert stellar frequencies to
determine the difference in structure between star and its model (see \citealt{earl2017} 
for recent attempts, successes and limitations), and we therefore need other means to judge the mismatch between the
internal structures of  stars and their models. One such way is to look at specific frequency combinations,
usually called separation ratios. The usual ones are
\be
r_{01}(n)=\frac{d\nu_{01}(n)}{\nu_{n,1}-\nu_{n-1,1}},\quad\quad 
r_{10}(n)=\frac{d\nu_{10}(n)}{\nu_{n+1,0}-\nu_{n,0}},
\label{eq:r01}
\ee
where,
\bea
d\nu_{01}(n)&=&\nu_{n,0}-\frac{1}{2}\left( \nu_{n-1,1}+\nu_{n,1} \right),\nonumber\\
d\nu_{10}(n)&=&\frac{1}{2}\left( \nu_{n,0}+\nu_{n+1,0} \right)-\nu_{n,1},
\label{eq:d01}
\eea
When plotted against frequency, $d\nu_{01}$ and $d\nu_{10}$ are not usually smooth and thus
sometimes higher order differences are used to define the separations (e.g., \citealt{iwr2009}, \citealt{victor2011},
\citealt{victor2015}). Another combination used extensively is
\be
r_{02}=\frac{\nu_{n+1,0}-\nu_{n,2}}{\nu_{n,1}-\nu_{n-1,1}}.
\label{eq:r02}
\ee
The ratio $r_{02}$ is sensitive to the central hydrogen concentration, and hence age,
of a star on the main sequence \citep{jcd1988}.

\begin{figure}
\includegraphics[width=3.00 true in]{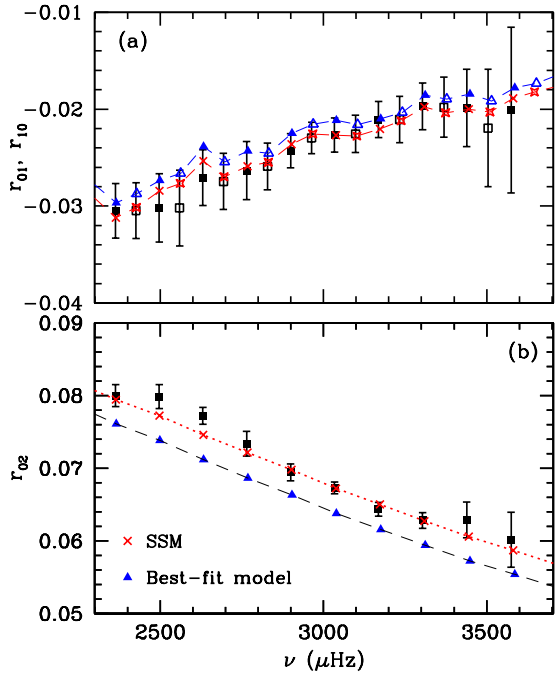}
\caption{The separation ratios $r_{01}$, $r_{10}$ (panel a) and $r_{02}$ (panel b) of
the Sun (points with error bars), the SSM (red crosses) and the best-fit model (blue triangles).
The ratios for the models have been connected with a line to guide the eye.
}
\label{fig:ratio}
\end{figure}

The separation ratios in Eq.~\ref{eq:r01}, and Eq.~\ref{eq:r02}\ are quite insensitive
to the surface term \citep[see][]{iwr2004, iwr2005, oti2005}.
In Fig.~\ref{fig:ratio} we compare the frequency ratios of the SSM and best-fit models
with those of the Sun, and can be seen clearly, the best-fit model does not really 
fit the ratios, in particular the mismatch is very large for $r_{02}$; this is of course
expected since $r_{02}$ is sensitive to the age of a star, and the
best-fit model has the wrong age. Thus had we used
the ratios to determine a best-fit model, we would not chosen this one. 

The insensitivity of the separation ratios to the surface term have led to the creation of
asteroseismic analysis pipeline that depend only on the separation ratios and not the
individual frequencies (see e.g., BASTA, \citealt{victor2015}) 
though the precision of results obtained with frequency rations appears to be smaller than those
obtained with frequencies explicitly corrected for the surface term
\citep[see e.g.,][]{victor2017}. One complicating factor in using the frequency
ratios is error-correlations. Errors for $r_{01}(n)$ and $r_{10}(n)$ are highly correlated,
as are errors between neighboring ratios of a given type,  e.g., between $r_{01}(n)$ and $r_{01}(n+1)$. 
Definitions of $r_{01}$ and $r_{10}$ that involve higher order differences are correlated
over many more data points.  The errors of neighboring $r_{02}$ are
also correlated. These correlations follow from the definition of the ratios.

\begin{figure}
\includegraphics[width=3.25 true in]{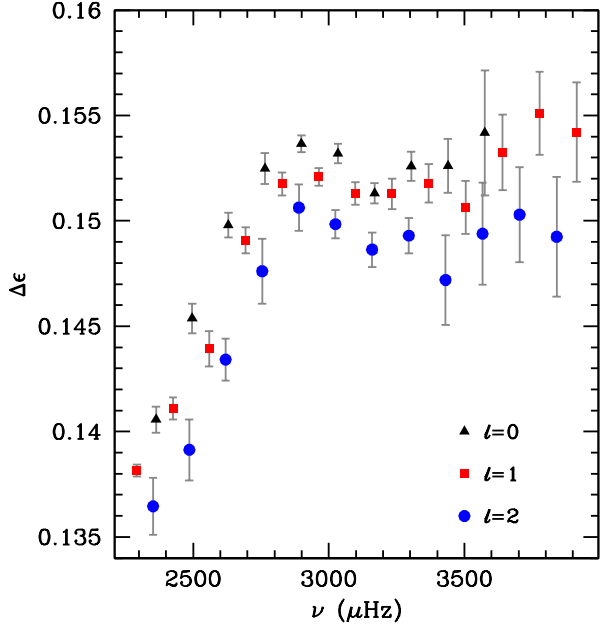}
\caption{The difference in \eps\ between Sun and the best-fit model. Note that at any given frequency,
the differences are degree dependent.
}
\label{fig:epsdif}
\end{figure}

Another quantity that can help in determining whether a model matches a star is the
phase factor \eps\ defined as:
\be
\nu=\langle\Delta\nu\rangle\left( n+\frac{l}{2}+\epsilon\right),
\label{eq:eps}
\ee
where $\langle\Delta\nu\rangle$ is the average large separation. It can be shown that
is the internal structure of a model matches that of a star, the difference in \eps\ between 
then is independent of degree and is a function of frequency alone \citep{iwr2015}.
 Fig.~\ref{fig:epsdif} shows the differences in \eps\ between the Sun and the
best-fit model; the differences are clearly degree dependent. \citet{iwr2015} demonstrated that \eps\ differences
can be used to determine the characteristics of a star.

Thus clearly the surface term correction can end up  giving us misleading results about a
star. It is worth noting, however, that while the best-fit model differs from the Sun in internal
structure, the global parameters of the model are not wildly different.
The mass of the best-fit models differs from that of the
Sun by only only 1.4\%, radius by 1\%, the difference in \teff\ is
negligible. The only major difference is in age, but that is only
a 6.5\% difference, much better than what is obtained by non-asteroseismic studies, even of
star clusters. Thus the question is whether this means that despite limitations of the
surface term, we can determine robust global properties of stars using their oscillation
frequencies. This is the question we try to answer in this paper. The global properties of
stars are what is use by the wider astronomy community for their work and hence, it is
important to examine the sensitivity of the results to the ubiquitous surface term.

We use all the three techniques --- explicit correction of the surface term using 
Eq.~\ref{eq:bl}, the use of frequency ratios, and the use of \eps\ differences to 
 examine whether the global properties
of the stars are indeed determined robustly despite the problem of the surface term.
We start with the Sun, and both components of the
16~Cyg system. These are three stars with the most well-determined oscillation frequencies. In order
to be sure that the quality of the data of these stars is not misleading us, we 
analyze three other stars, KIC~6106415 (HD~177153), KIC~6225718 (HD~187637), and KIC~8006161 (HIP~91949).

\section{The analysis techniques}
\label{sec:method}

We estimate the global properties of the stars we are studying by  statistical analyses of a 
large ensemble of models. We determine the likelihood of each model and the final parameters
are the likelihood-weighted average of the underlying properties of the models. What differs
in each case what quantity is used to define the likelihood. Since we do not have evidence
to the contrary, all inputs are assumed to have a Gaussian distribution of errors that allows
us to define a $\chi^2$ value for each input.

We start with the method of \eps\ differences. For each model we calculate the difference in \eps\ between
the star and the model, with the differences being calculated at the observed frequencies, i.e.,
\be
\Delta\epsilon_{nl}=\epsilon^{\rm obs}(\nu^{\rm obs}_{nl})-\epsilon^{\rm model}(\nu^{\rm obs}_{nl}),
\label{eq:epdif}
\ee
where, $\epsilon^{\rm model}(\nu^{\rm obs}_{nl})$ is \eps\ for the model at degree $l$ interpolated
to the observed frequency $\nu^{\rm obs}_{nl}$ of the same degree $l$.
To determine
whether or not the \eps\ differences are a function of frequency alone, we define an arbitrary 
function of frequency ${\mathcal F}=\sum_i^Ma_i\phi(\nu)_i$, where $\phi(\nu)$ are basis
functions in frequency. We then minimize
\be
\chi^2(\epsilon)=
\frac{1}{N-M}\sum_{nl}\left(\frac{\Delta\epsilon_{nl}-{\mathcal F}(\nu^{\rm obs}_{nl})}{S^{\rm obs}_{nl}}
\right),
\label{eq:chiep}
\ee
where $N$ is the total number of models, $M$ the number of basis function (chosen to be fewer than the
number of radial modes as per the suggestion of \citealt{iwr2015}), and 
$S^{\rm obs}_{nl}=\sigma^{\rm obs}_{nl}/\Delta\nu^{\rm obs}$, with $\sigma^{\rm obs}_{nl}$ being the
uncertainty on $\nu^{\rm obs}_{nl}$. We use B-spline basis functions. A small value of
$\chi^2(\epsilon)$ indicates that the \eps\ differences are a function of frequency alone. We then define the
likelihood of any given model as
\be
{\mathcal L}(\epsilon)=A\exp\left(-\frac{\chi^2(\epsilon)}{2}\right),
\label{eq:epslike}
\ee
$A$ being a normalization factor such that $\sum{\mathcal L}(\epsilon)=1$ with the sum taken over all models.

Next, we consider frequency ratios. To avoid large error-correlations we only use ratios $r_{01}$ and $r_{02}$
as define in Eqs.~\ref{eq:r01} and \ref{eq:r02}. We do not use $r_{10}$, and we neglect the error correlations
between $r_{01}$ and $r_{02}$. We then define the likelihood for $r_{01}$ as
\be
\chi^2(r_{01})=({\overbar{r_{01}}}^{({\rm obs})}-{\overbar{r_{01}}}^{({\rm model})})^T{\mathbf C}^{-1}
({\overbar{r_{01}}}^{({\rm obs})}-{\overbar{r_{01}}}^{({\rm model})}),
\label{eq:xhir01}
\ee
where ${\overbar{r_{01}}}({\rm obs})$ is the vector defining the observe $r_{01}$,
${\overbar{r_{01}}}({\rm model})$ is the vector defining the $r_{01}$ for the
model at the observed frequency, and ${\mathbf C}$ is the error-covariance matrix.
Thus ${\mathcal L}(r_{01})=B\exp(-\chi^2(r_{01})/2)$.
We define ${\mathcal L}(r_{02})$ in an analogous manner, and thus for the two ratios taken
together
\be
{\mathcal L}({\rm rat})={\mathcal L}(r_{01}){\mathcal L}(r_{02}).
\label{eq:ratlike}
\ee

The third way of analyzing the models was to use surface-term corrected frequencies
of the models. We define $\nu^{\rm corr}_{nl}=\nu_{nl}^{\rm model}-S$, where $S$ is
defined by the right-hand side of Eq.~\ref{eq:bl}. We can then define 
\be
\chi^2(\nu)=\frac{1}{N-2}\sum_{nl}\frac{(\nu_{nl}^{\rm obs}-\nu_{nl}^{\rm corr})^2}{\sigma^{\rm obs}_{nl}},
\label{eq:eqchinu}
\ee
and consequently
\be
{\mathcal L}(\nu)=C\exp\left(-\frac{\chi^2(\nu)}{2}\right),
\label{eq:nulike}
\ee
$C$ being the normalization constant.

The surface term is expected to be smaller at low frequencies and larger at high frequencies,
however, for models that do not fit the data very well, the frequency differences can show the 
opposite behavior.  We weigh against such models by defining a weight ${\mathcal W}({\rm low})$
where
\be
{\mathcal W}_{\rm low}=\exp\left(-\frac{\chi^2_{\rm low}}{2}\right),
\label{eq:wlow}
\ee
where $\chi^2_{\rm low}$ is the $\chi^2$ for the (uncorrected) frequency differences
between the two lowest frequency radial modes.

It is usual to consider the observed \teff\ and metallicity in the analyses, and we
have analyzed the models with and without constraints on \teff\ and metallicity. The
likelihood for \teff\ is defined as
\be
{\mathcal L}(T)=D\exp(-\chi^2(T)/2),
\label{eq:tcal}
\ee
with
\be
\chi^2(T)=\frac{(T^{\rm obs}_{\rm eff}-T^{\rm model}_{\rm eff})^2}{\sigma^2_{ T}},
\label{eq:chit}
\ee
where $\sigma_{T}$ is the uncertainty on the effective temperature, we consider
two cases, $\sigma_{T}=100$K and $\sigma_{\rm T}=50$. We similarly define a
likelihood ${\mathcal L}({\mathrm{[Fe/H]}})$ and again we consider two different
uncertainties, $\sigma_{\mathrm{[Fe/H]}}=0.1$ dex and $\sigma_{\mathrm{[Fe/H]}}=0.05$ dex.

For some  cases we also apply a weight for age.
The equations of stellar structure and evolution do not know anything about the age of the
universe, and thus in principle, a model could gave an age greater than that of the universe.
While a  model older than the age of the universe is unphysical, uncertainties
in mass, metallicity and effective temperature could easily result in  such a
model. Instead of removing all models above an age of 13.8 Gyr (which would
result in a sharp cut-off in the probability density function of age) we use a weight for age
$\tau$ defined as
\be
{\mathcal W}_{\rm age}=
\begin{cases}
1, & \hbox{if}\;\; \tau <=13.8\;\hbox{Gyr}\\
\exp\left[-\frac{(13.8-\tau)^2}{2\sigma^2_\tau}\right] & \hbox{otherwise},\\
\end{cases}
\label{eq:wage}
\ee
$\tau$ is in units of Gigayears, and  $\sigma_{\tau}$ is chosen to be 0.1 Gyr.

In Table~\ref{tab:methods} we list all the different parameter combinations that we 
have used to determine the global properties of the stars under question.
Method~1 simply means that we calculate the likelihood of a model only with \eps, 
i.e., we only calculate ${\mathcal L}({\rm total})={\mathcal L}(\epsilon)$, in Method~2, it is calculated
as ${\mathcal L}({\rm total})={\mathcal W}_{\rm age}{\mathcal L}(\epsilon)$, in Method~3 corresponds to
${\mathcal L}({\rm total})={\mathcal L}(\epsilon){\mathcal L}(T){\mathcal L}(\hbox{[Fe/H]})$, etc.

\begin{deluxetable*}{cll}
\tablecolumns{3}
\tablecaption{The different combinations of observables used in the analysis}
\tablehead{\colhead{Method No.}& \colhead{Observables used} &\colhead{Notes}\label{tab:methods}}
\startdata
\phantom{1}1 & $\epsilon$ &\\
\phantom{1}2 & $\epsilon$ & additional weights for age using ${\mathcal W}_{\rm age}$\\
\phantom{1}3 & $\epsilon$, \teff, \feh  & using $\sigma_{ T}=100$K and  $\sigma_{\mathrm{[Fe/H]}}=0.1$ dex\\
\phantom{1}4 & $\epsilon$, \teff, \feh  & using $\sigma_{ T}=75$K and  $\sigma_{\mathrm{[Fe/H]}}=0.05$ dex\\
\phantom{1}5 & $\epsilon$, \teff, \feh  & as in Method 4, but weighted for age using ${\mathcal W}_{\rm age}$\\
\phantom{1}6 & $r_{01},\;r_{02}$ &\\
\phantom{1}7 & $r_{01},\;r_{02}$ & additional weights for age using ${\mathcal W}_{\rm age}$\\
\phantom{1}8 & $r_{01},\;r_{02}$, \teff, \feh  & using $\sigma_{ T}=100$K and  $\sigma_{\mathrm{[Fe/H]}}=0.1$ dex\\
\phantom{1}9 & $r_{01},\;r_{02}$, \teff, \feh  & using $\sigma_{ T}=75$K and  $\sigma_{\mathrm{[Fe/H]}}=0.05$ dex\\
         10  & $r_{01},\;r_{02}$, \teff, \feh  & as in Method 9, but weighted for age using ${\mathcal W}_{\rm age}$\\
         11  & $\nu$ &\\
         12  & $\nu$ & additional weights for age using ${\mathcal W}_{\rm age}$\\
         13  & $\nu$, \teff, \feh  & using $\sigma_{ T}=100$K and  $\sigma_{\mathrm{[Fe/H]}}=0.1$ dex\\
         14  & $\nu$, \teff, \feh  & using $\sigma_{ T}=75$K and  $\sigma_{\mathrm{[Fe/H]}}=0.05$ dex\\
         15  & $\nu$, \teff, \feh  & as in Method 14, but weighted for age using ${\mathcal W}_{\rm age}$\\
         16  & $\nu$, \teff, \feh  & as in Method 15, but with additional weight ${\mathcal W}_{\rm low}$\\
\enddata
\end{deluxetable*}

\section{The data}
\label{sec:stars}

\begin{deluxetable*}{lcccc}
\tablecolumns{5}
\tablecaption{Average seismic parameters and surface properties\label{tab:data}}
\tablehead{\colhead{Star} & \colhead{\dnu} & \colhead{\numax} & \colhead{\teff} &\colhead{\feh}\\
   & \colhead{($\mu$ Hz)} & \colhead{($\mu$ Hz)} & \colhead{(K)} & }
\startdata
Sun         & $134.91 \pm 0.02$ & $3073 \pm 13$ & $5772 \pm 10$ & $0.0 \pm 0.0$\\
16~Cyg~A    & $103.28 \pm 0.02$ & $2188 \pm 5 $ & $5825 \pm 50$ & $0.01\pm 0.03$\\
16~Cyg~B    & $116.93 \pm 0.02$ & $2561 \pm 5 $ & $5750\pm 50$ & $0.05 \pm 0.02$ \\
KIC~6106415 & $104.07 \pm 0.03$ & $2249 \pm 5 $ & $6037 \pm 77 $& $ -0.04\pm 0.1$\\
KIC~6225718 & $105.07 \pm 0.02$ & $2364 \pm 5 $ & $6313 \pm 77$ & $ -0.07\pm 0.1$\\
KIC~8006161 & $149.43 \pm 0.02$ & $3575 \pm 11$ & $5488 \pm 77$ & $ 0.34 \pm 0.1$\\
\enddata
\end{deluxetable*}

We analyze six stars in this paper. The first three, the Sun, 16~Cyg A and 16~Cyg B not only have precise
data on frequencies, they also have precise data on their effective temperatures and metallicities. 
For the Sun, we use the degraded solar frequencies used by \citet{victor2017} as obtained
by \citet{lund2017}. For each star, 
we use frequencies, large separations and \numax\ values from \citet{lund2017}, although for \dnu\ and \numax, we
assume that the uncertainty is the larger of the two limits provided.
We analyze three others stars: KIC~6106415 (HD~177153), 
KIC~6225718 (HD~187637), and KIC~8006161 (HIP~91949). 
The frequencies of these stars are not as precise as those of the others, and the \teff\ and \feh\ values
are not known as precisely as those of the other stars either.
The
results for these stars allow us to judge the dependence of our results on the quality of the data. 
For the Sun, we assume the \teff\ value of \citet{mamajek2015}. For the components of the
16~Cyg system we use \teff\ and \feh\ from \citet{ramirez2009}, KIC~6106415, KIC~6225718 and
KIC~8006161 we use \feh\ from \citet{lars2015} and temperatures derived by Buchhave \& Latham as listed in
\citet{savita2017}.
Table~\ref{tab:data} lists the average seismic properties, \teff\ and \feh\ of the stars being studied.

\section{Construction of stellar models}
\label{sec:models}

We constructed a large number of models for each star with YREC, the Yale
stellar evolution code \citep{demarque2008}. The models were constructed with the OPAL equation of state
\citep{eos}. OPAL opacities \citep{opac} supplemented with low-temperature opacities from \citet{ferg} were used.
We used the NACRE reaction rates from \citet{nacre} except for the $^{14}N(p,\gamma)^{15}O$ reaction,
where we used the rates of \citet{marta}. We included diffusion and settling of heavy elements 
using the coefficients of \citet{thoul}. To avoid the problem of heavy-elements draining out of the
outer convection zones of hot stars in an unphysical manner, we artificially reduced the diffusion 
coefficients as a function of total  mass as:
\be
C_{\rm diff}=
\begin{cases}
1, & \hbox{if}\;\; M <=1.25\\
\exp\left[-\frac{(M-1.25)^2}{2(0.085)^2}\right] & \hbox{otherwise},\\
\end{cases}
\label{eq:diffcoef}
\ee
where $M$ is the mass of the model in units of \msun, and $C_{\rm diff}$ a multiplicative
factor for the \citet{thoul} diffusion coefficients.

\begin{deluxetable}{lcc}
\tablecolumns{3}
\tablecaption{Range of \teff\ and initial \feh\ of models\label{tab:range}}
\tablehead{\colhead{Star} & \colhead{\teff} &\colhead{\feh}\\
& \colhead{(K)} & }
\startdata
Sun         & 5550--5950 & $-0.33$--$0.47$ \\
16~Cyg~A    & 5640--6040 & $-0.20$--$0.60$ \\
16~Cyg~B    & 5610--6010 & $-0.25$--$0.55$ \\
KIC~6106415 & 5730--6345 & $-0.30$--$0.40$ \\
KIC~6225718 & 6005--6620 & $-0.10$--$0.70$ \\
KIC~8006161 & 5180--5800 & $\phantom{-}0.10$--$0.80$\\
\enddata
\end{deluxetable}

We start the modelling process with the observed \dnu, \numax, \teff, and \feh\ of each star,
along with their uncertainties. Using these we created many realizations of the four quantities.
For \dnu, and \numax, we assumed that the realizations have a Gaussian distribution with a $\sigma$ of
four times the uncertainty in each quantity. We assumed flat distributions  for \teff and \feh.
Since the observed \feh\ is the present day metallicity, we assumed that the
initial metallicity was about 0.1~dex higher than the present one for the purpose of constructing models.
The spread in current \teff\ and initial \feh\ for models of each star are listed in Table~\ref{tab:range}.
For the initial helium abundance $Y_0$ we assumed a flat distribution between $0.24$ and $0.34$.
The initial \feh\ was converted to initial hydrogen abundance $X_0$ and heavy-element abundance $Z_0$ using
the initial $Y_0$ assuming the \citet{gs98} solar  metallicity scale, i.e., $\hbox{[Fe/H]}=0\equiv Z/X=0.023$.

\begin{figure}
\includegraphics[width=2.50 true in]{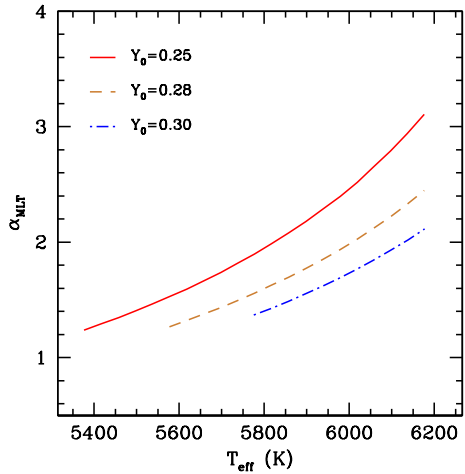}
\caption{The mixing-length parameters $\alpha_{\rm MLT}$ required for a 1\msun, \feh=0 model to have
a radius of 1\rsun\ at different values of \teff. We show the case for three different values of
$Y_0$. Unlike rest of the models used in this work, models in this figure were constructed
without diffusion and settling of heavy elements to ensure that the models have the
 exact value of the current surface metallicity.
}
\label{fig:alpha}
\end{figure}

For any given realization, we have a set of parameters
(\dnu, \numax, \teff, $X_0$, $Z_0$, $Y_0$). For each set, \dnu\ and \numax\ were converted to 
mass $M$ and radius $R$ using the asteroseismic scaling relations corrected as per the prescription of 
\citet{guggenberger2016}. This gave us sets of ($M$, $R$, \teff, $X_0$, $Z_0$), where
$M$, $X_0$, $Z_0$ are inputs to the models and we want a model of radius $R$ at the specified  \teff\ to be the output.
This is done by using YREC in an iterative manner where we iterate over $\alpha_{\rm MLT}$, the
mixing length parameter, until we obtain a models that has the required radius at the specified
\teff. The condition on the observed present-day metallicity is applied {\it post facto} when we use
Methods~3--5, 8--10, and 13-16 to analyze the ensemble of models; it is ignored otherwise.
In Fig.~\ref{fig:alpha}\ we show an example of how $\alpha_{\rm MLT}$ has to change with \teff\ in order
for a 1\msun\ model to have a radius of 1\rsun\ at that value of  \teff.

The frequencies of each model were calculated using the code of \citet{ab94} and 
were then used to calculate \eps, and frequency ratios $r_{01}$ and $r_{02}$. For models of any
given star, we only used those modes that have been observed. The frequency difference of each model
with respect to the observed frequencies were fitted to the Ball \& Gizon form (Eq.~\ref{eq:bl})
to determine the surface-term corrected frequencies. The ensemble of models of each star were
then analyzed using different measures as listed in Table~\ref{tab:methods}.

\section{Results}
\label{sec:res}

\begin{figure*}
\centerline{
\includegraphics[width=2.25 true in]{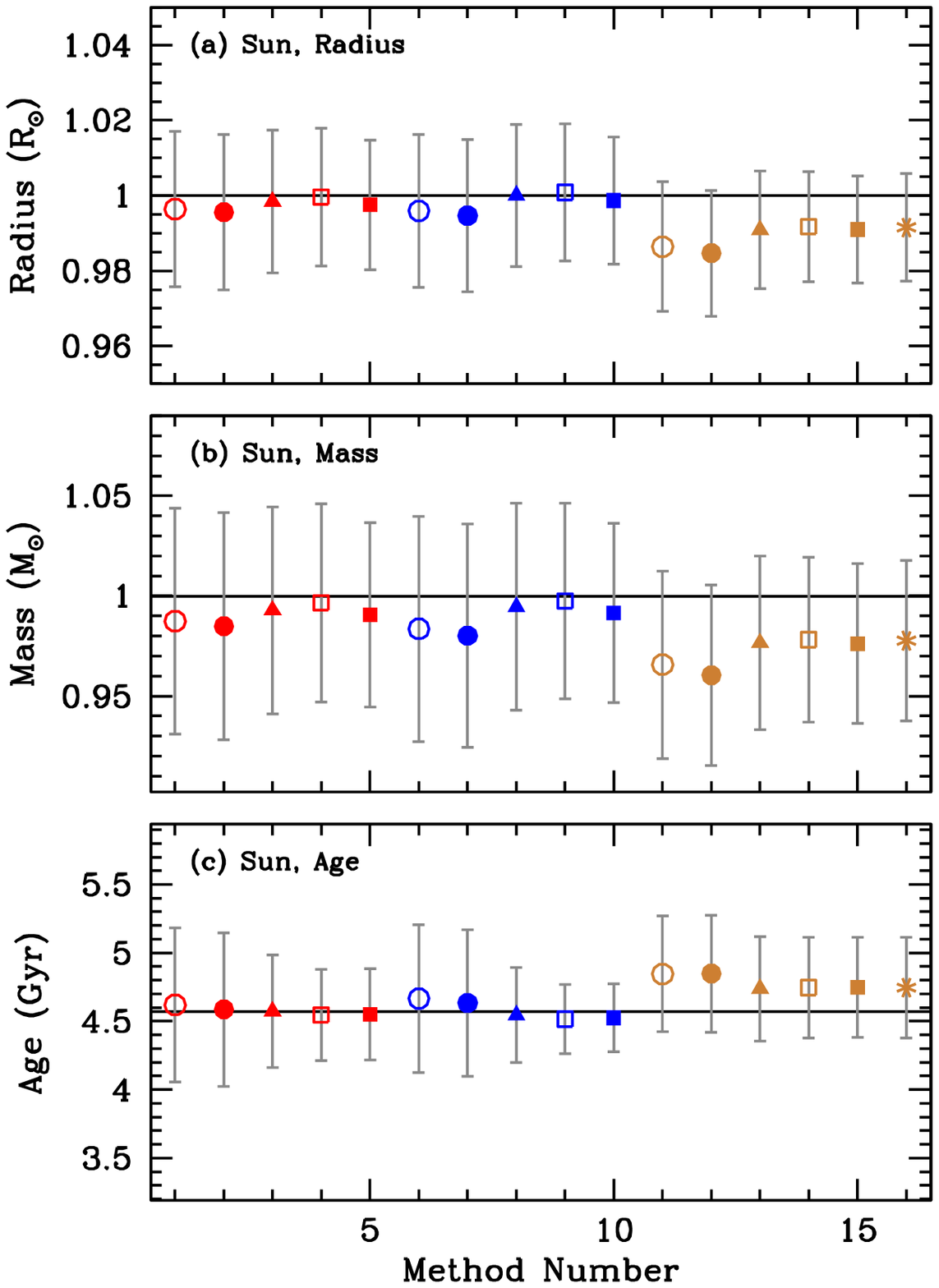}
\includegraphics[width=2.25 true in]{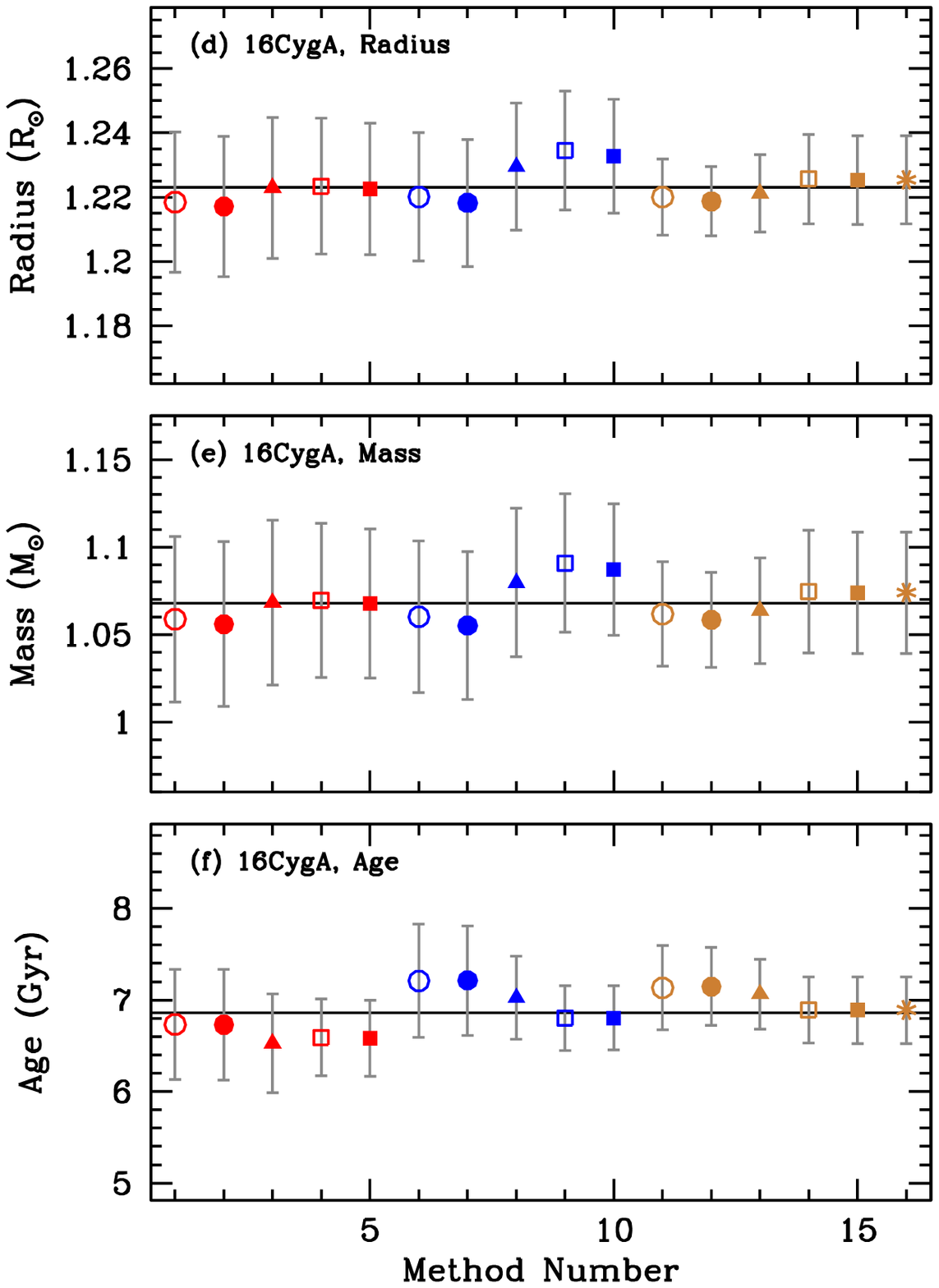}
\includegraphics[width=2.25 true in]{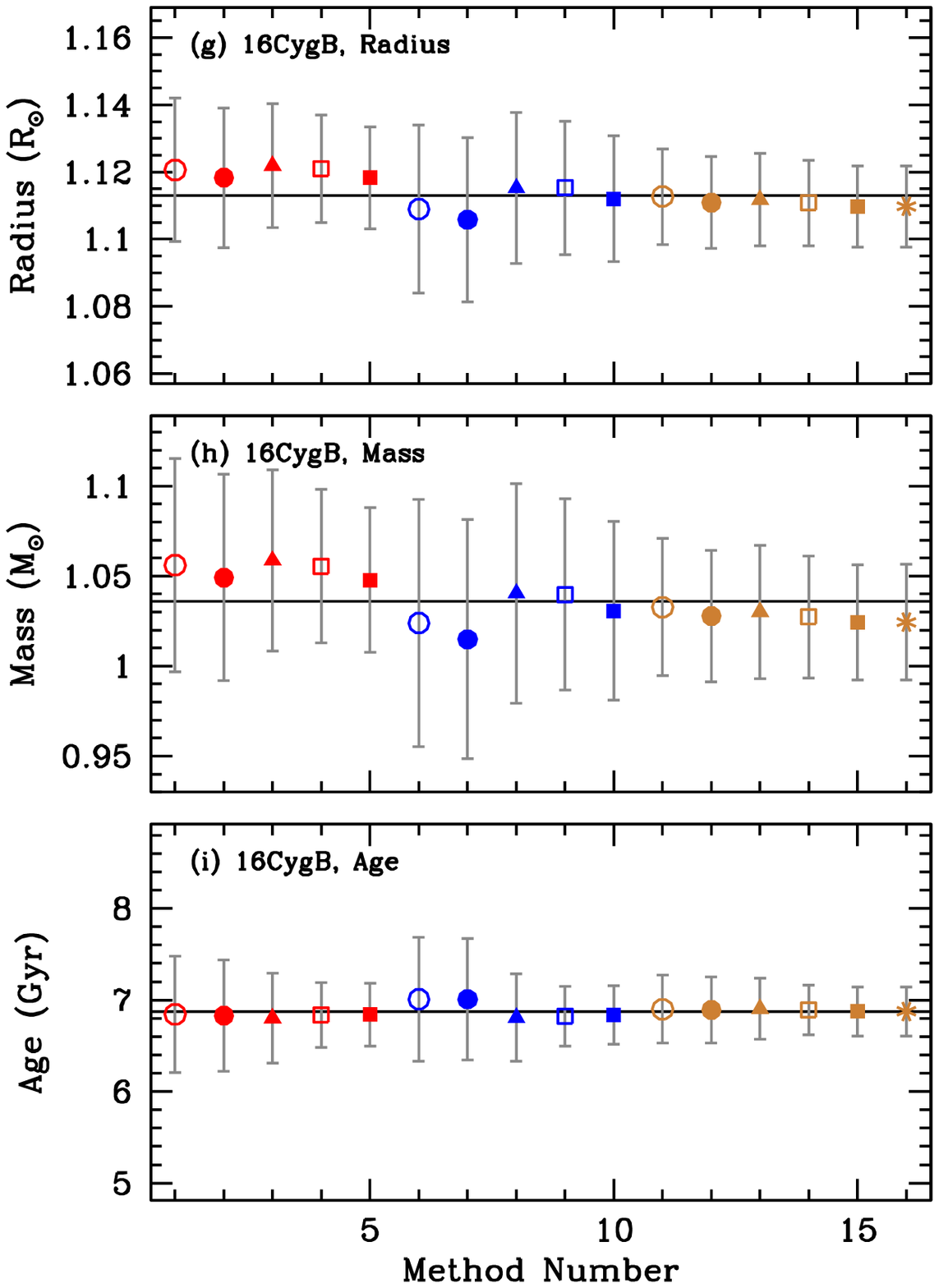}
}
\caption{Estimates of mass, radius and age of the Sun, 16 Cyg~A and 16 Cyg~B obtained using
goodness-of-fit measures to different observables as listed in Table~\ref{tab:methods}. The
numbers on the abscissa refer to the method number in the table. Red, blue and orange
points denote the fact that the seismic variable used is \eps, frequency ratios and
surface-term corrected frequencies respectively. The horizontal lines for the solar case are
the known solar values. For the other cases we plot the mean of the results obtained. The vertical
extent of the panels shows a fixed fraction of the values ---  $\pm 5$\%  for radius, 
$\pm 10$\% for mass, and $\pm 30$\% for age --- to give an indication of the precision of the
estimates.
}
\label{fig:rma1}
\end{figure*}

The mass, radius and age of the Sun, and the two components of the 16~Cyg system as derived
using the sixteen different combinations are shown in Fig.~\ref{fig:rma1}. It is very clear from the
figure that central value of the estimates (i.e., the likelihood weighted means) are very
similar in each case. The uncertainty on the results on the other hand, are somewhat different. The methods
that explicitly use the surface-term corrected frequencies appear to have the lowest uncertainties,
however, at least for the Sun where we have independent measures of mass and radius, the
surface-term corrected frequencies give the largest systematic errors. But in all cases, the results
are well within 1$\sigma$ of the actual value. Unlike radius and mass, age uncertainties are lowered
considerably
when \teff\ and \feh\ are considered explicitly --- not surprising since the age of a model at a
given mass and radius depends critically on its temperature and metallicity. 
It should be noted that all results presented here are consistent with those of
\citet{victor2017}.

\begin{figure*}
\centerline{
\includegraphics[width=2.25 true in]{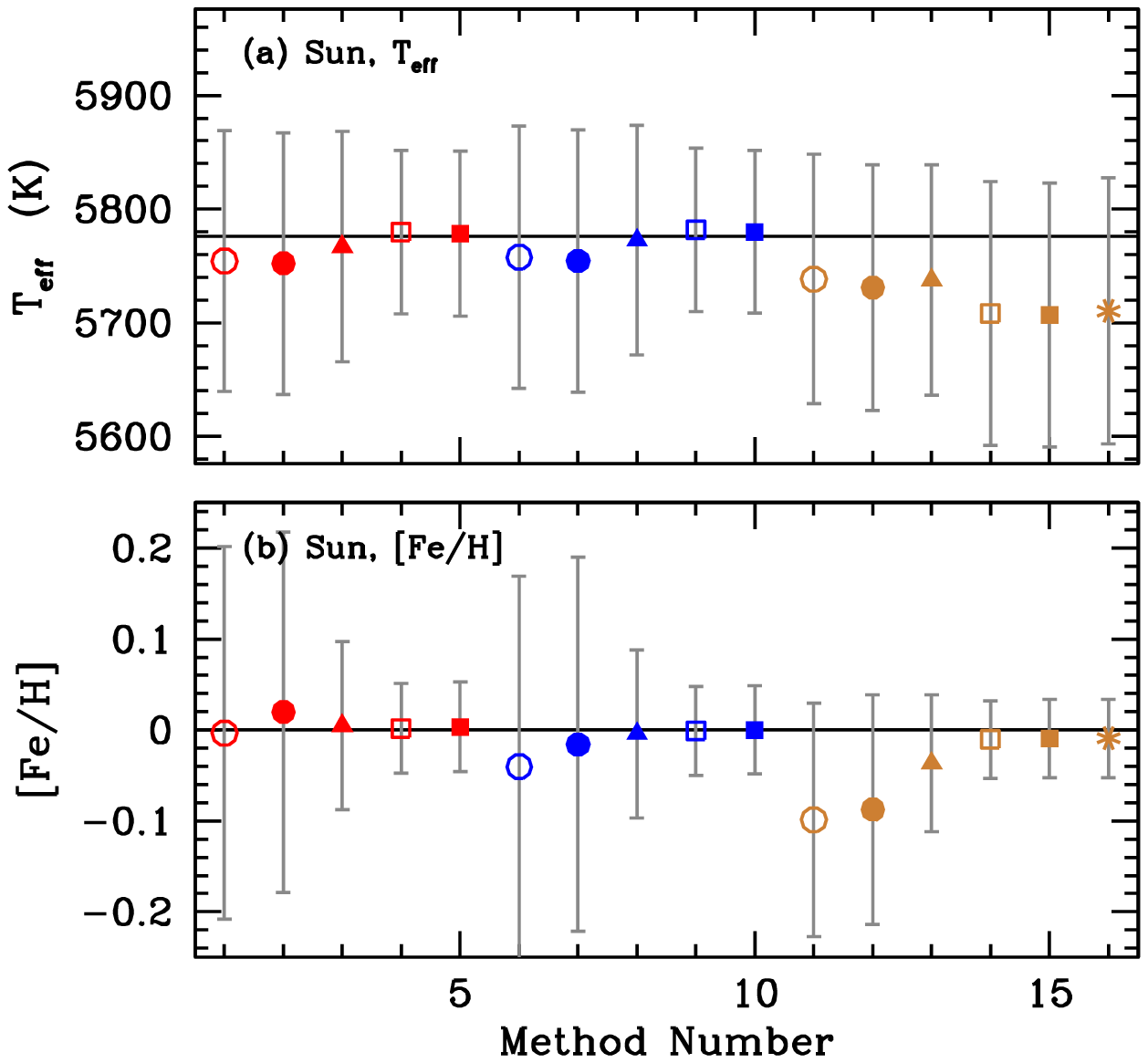}
\includegraphics[width=2.25 true in]{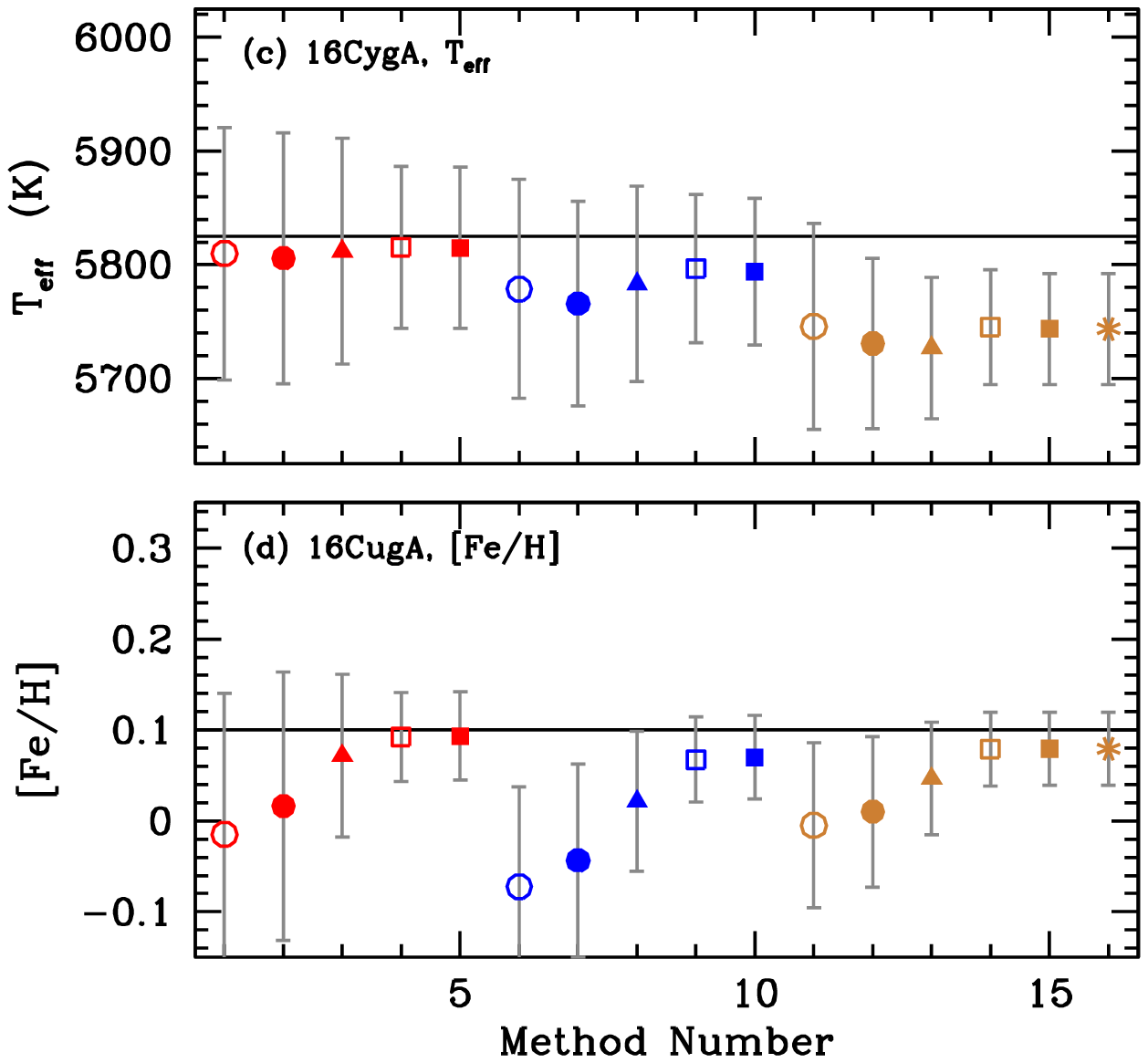}
\includegraphics[width=2.25 true in]{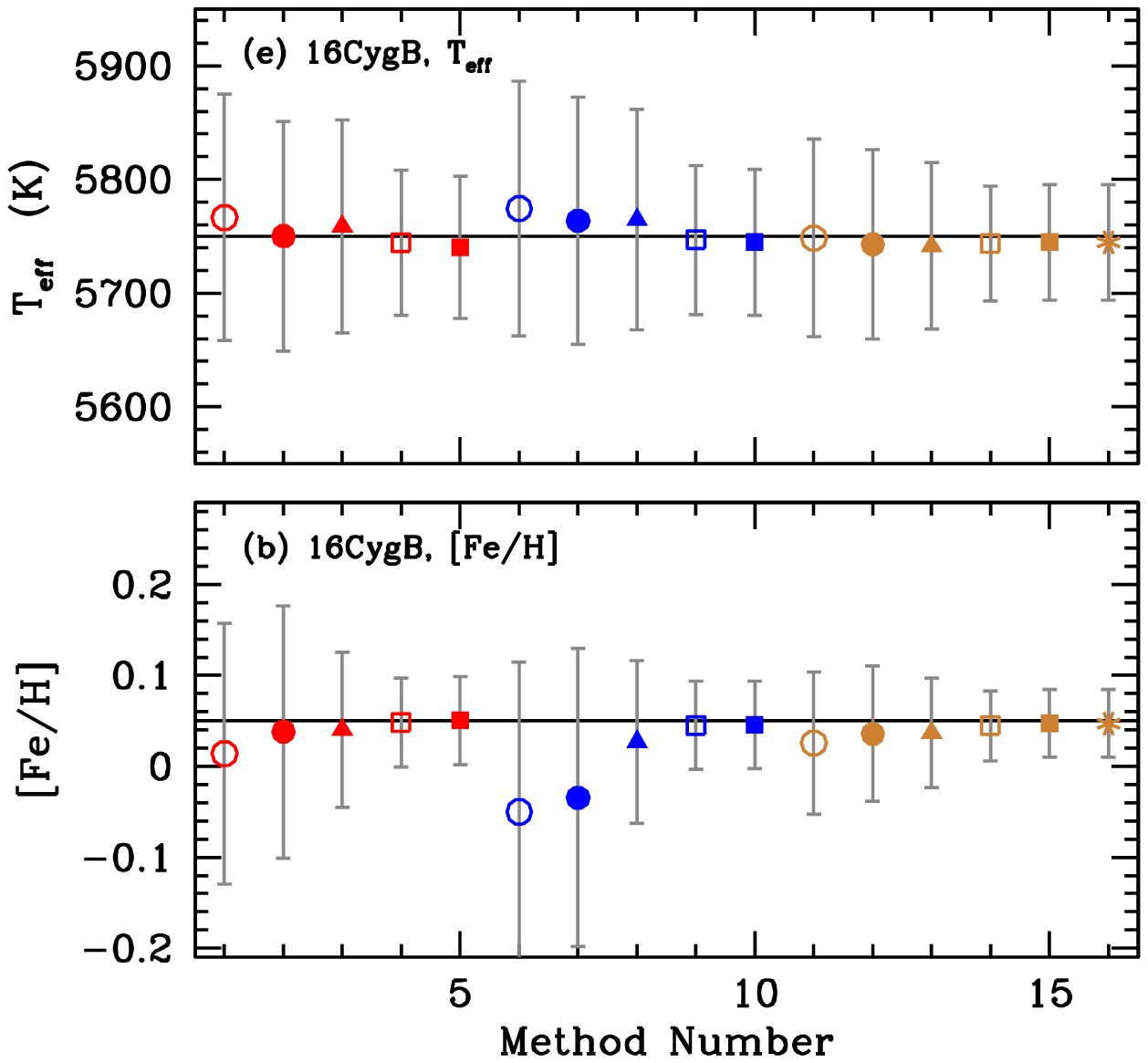}
}
\caption{\teff\ and \feh\ for the Sun, 16 Cyg~A and 16 Cyg~B. In the panels that show results for
the Sun, the horizontal line marks the known solar value; in the other panels the line is
simply an arithmetic average of all the results shown in the panel.
}
\label{fig:tz}
\end{figure*}

Asteroseismic analyses use \teff\ and \feh\ to complement the data set, however, we 
attempted to examine how well we could estimate the temperature and metallicity
of stars from seismic data alone. The results for the Sun and the 16 Cyg system are
shown in Fig.~\ref{fig:tz}. We find that temperature can be localized to better than
$\pm 100$K when no constraint is applied, when we do apply a temperature constraint, we
find that we recover temperature to about $\pm 70$K regardless of whether we assumed
a Gaussian spread of 100~K or 75~K. We should note however, that although we
assumed a flat prior in temperature, the models were restricted to a finite (400~K) range in
temperature, and that could have an effect on the final spread of the results. Metallicity results are worse;
 unless we apply a constraint,
we can recover \feh\ to only between 0.1 and 0.2~dex. We get more precise results when the 
explicit surface-term corrections are used, but judging by the solar case, that can result in larger  systematic
errors.

\begin{figure}
\includegraphics[width=2.75 true in]{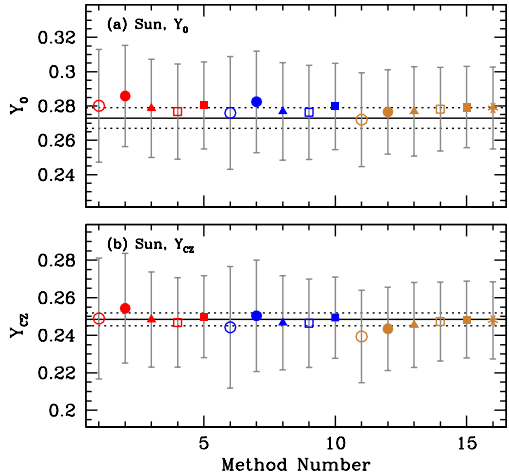}
\caption{The initial helium abundance and the current convection-zone helium 
abundance of the Sun. The solid horizontal line in each panel marks the
helioseismically determine value of that quantity --- $Y_0$ from \citet{aldo2010}
and $Y_{\rm CZ}$ from \citet{basu1998}. 
Dotted lines show 1$\sigma$ uncertainties of the helioseismic estimates.
}
\label{fig:ycz}
\end{figure}
\begin{figure}
\includegraphics[width=2.75 true in]{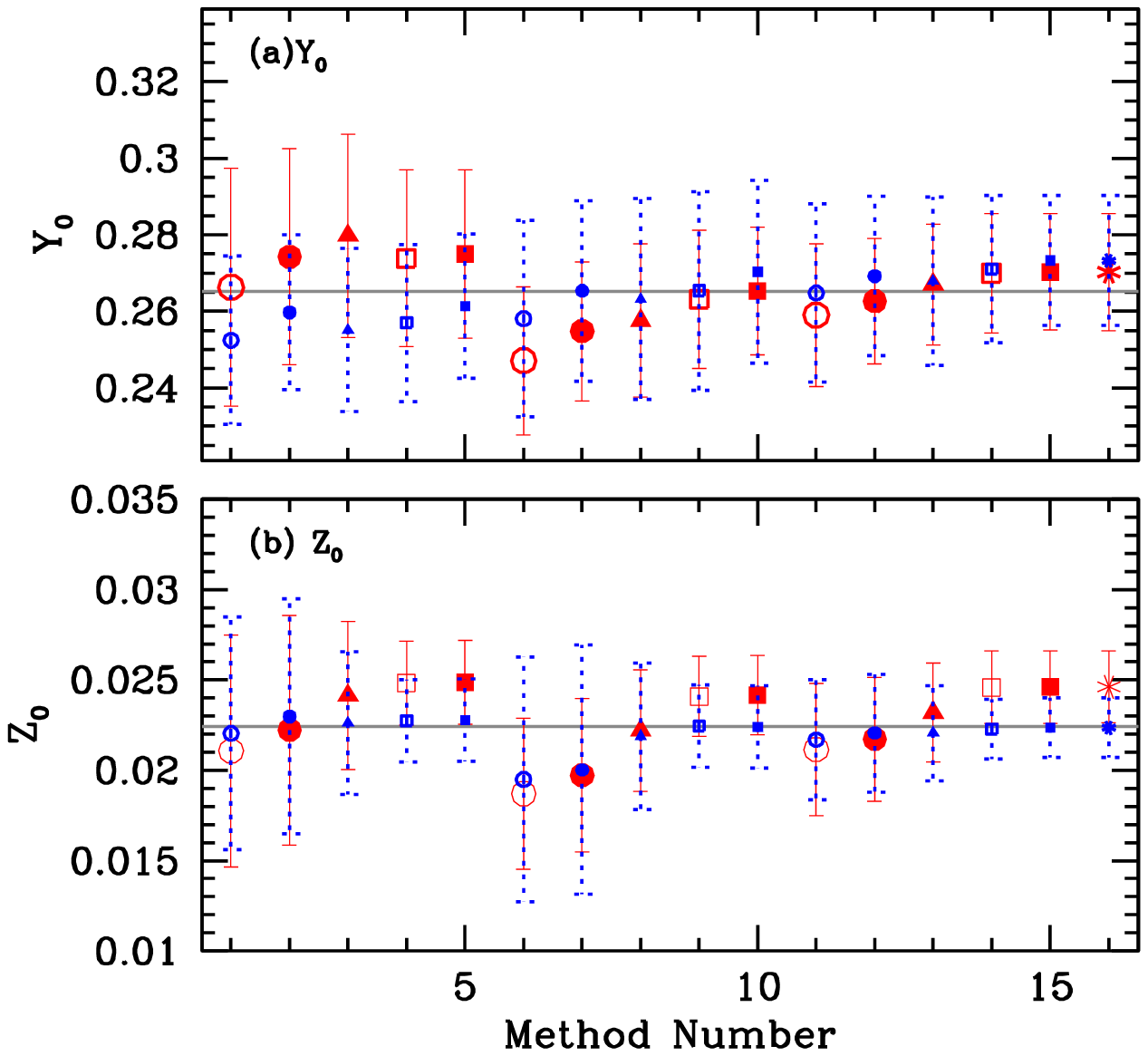}
\caption{The initial helium and heavy-element abundance of 16 Cyg A (large red points)
and 16 Cyg B (small blue points). 
}
\label{fig:cygy}
\end{figure}

For the Sun, we find that no matter which method we use, we can reproduce the helioseismically
determined initial helium abundance of the Sun \citep{aldo2010} as well the current 
convection-zone helium abundance for the Sun \citep[see][]{basu1998}. These are shown in Fig.~\ref{fig:ycz}.
The precision for $Y_0$ is however, a factor of five worse than that obtained
from the detailed helioseismic study. The uncertainty in $Y_{\rm CZ}$ is a factor of ten worse. However,
the difference between $Y_0$ and $Y_{\rm CZ}$ does give a good estimate of the amount of helium
that has settled out of the convection zone. The analysis of the two components of the 16~Cyg system
shows that estimates of $Y_0$ and $Z_0$ are the same for both components of the system (Fig.~\ref{fig:cygy})
 even though we did not take into account the fact that the two stars should have the
same initial composition.

\begin{figure*}
\centerline{
\includegraphics[width=2.25 true in]{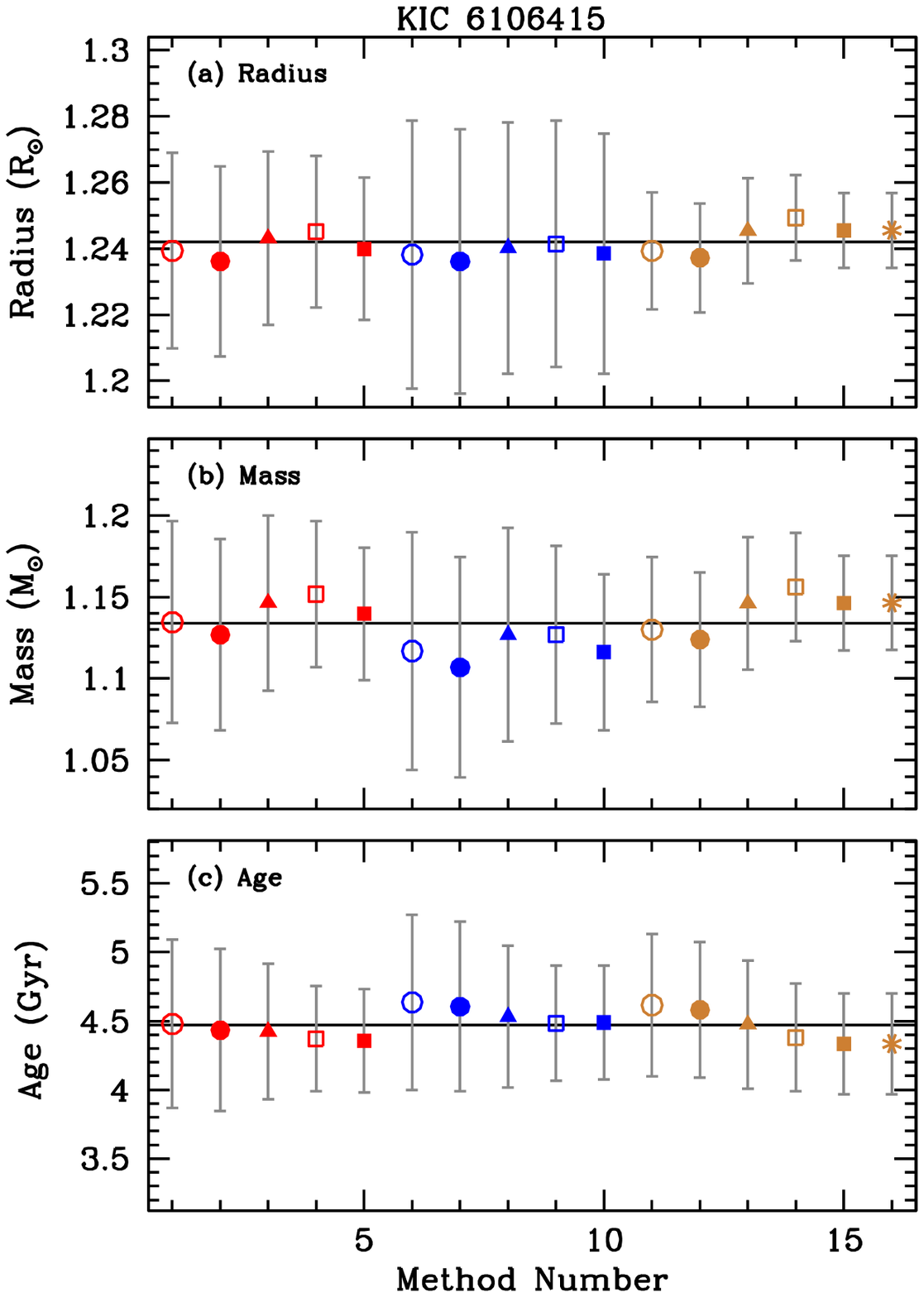}
\includegraphics[width=2.25 true in]{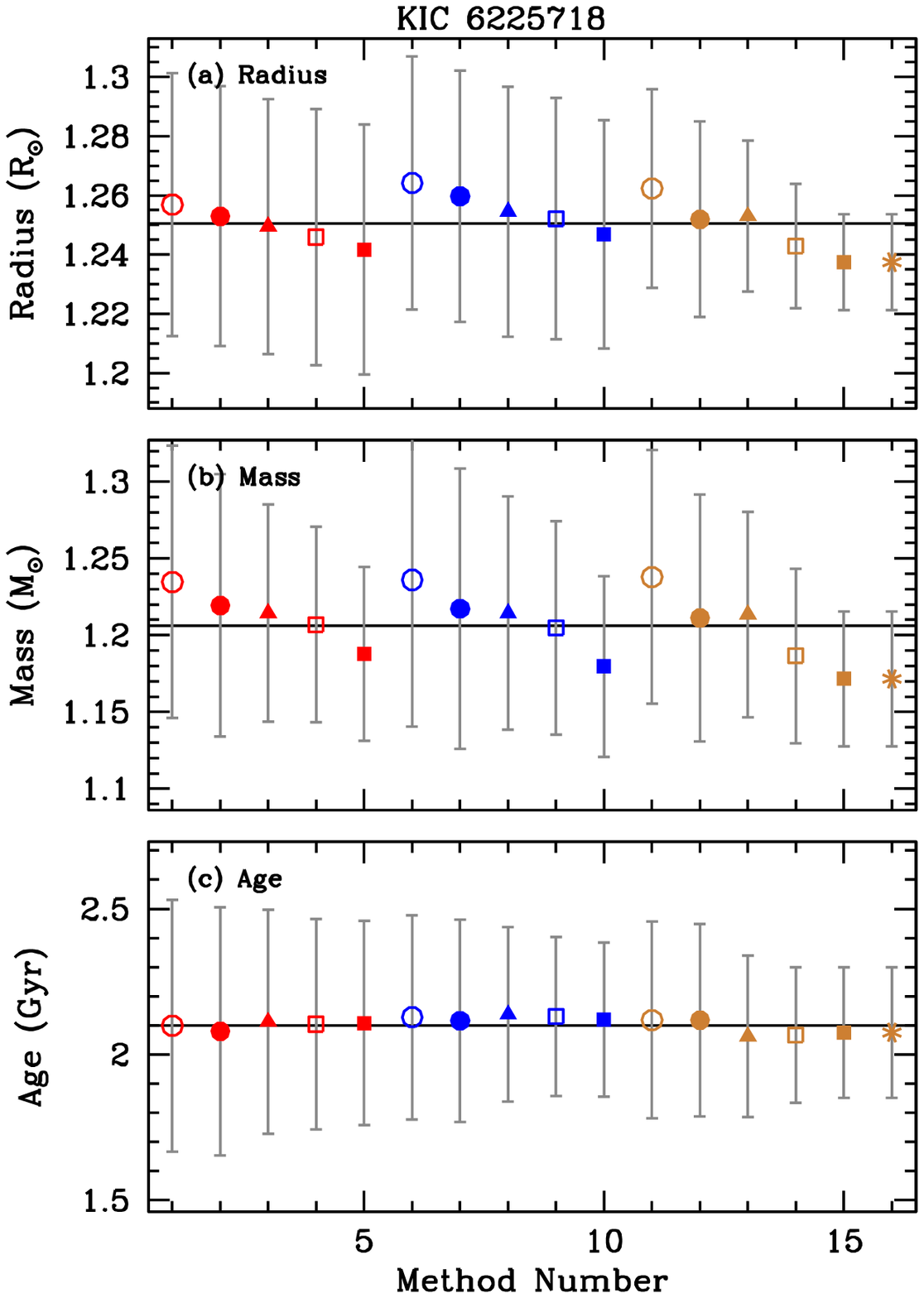}
\includegraphics[width=2.25 true in]{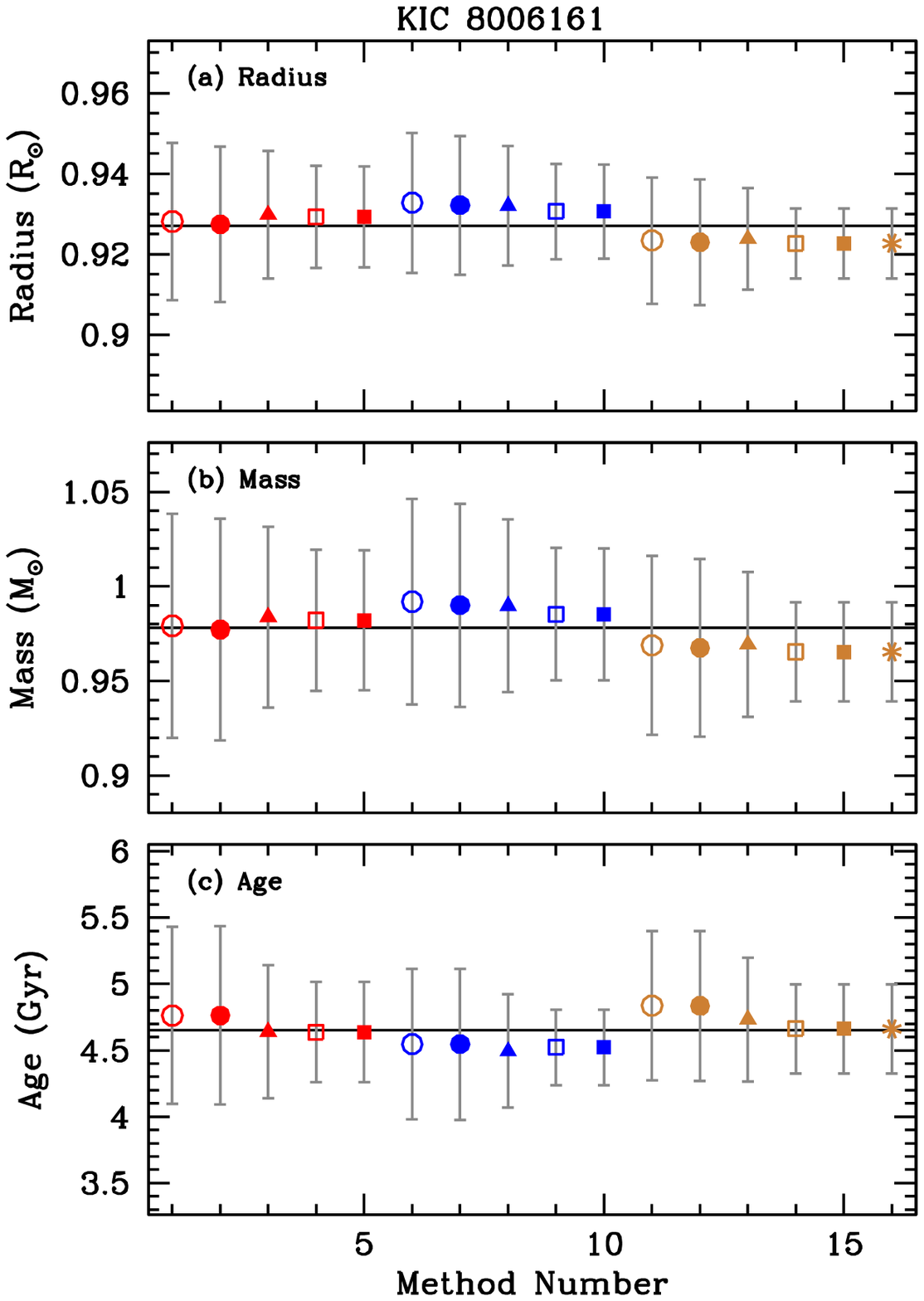}
}
\caption{Mass, radius and age estimates for KIC~6106415, KIC~6225718 and KIC~8006161.
The colors are the same as in Fig.~\ref{fig:rma1} The 
horizontal line in each panel is the average of all the values in that panel.
}
\label{fig:rma2}
\end{figure*}

The Sun and the two stellar components of the 16~Cyg system have the most precisely determined
frequencies, and this leads to the question of what happens if the frequency-set
is not as extensive, or if the errors on the modes larger. To answer this we analyzed the
data of the other three stars and the results are shown in Fig.~\ref{fig:rma2}. As as be
seen, the results obtained for a given star with using the different asteroseismic parameters
are consistent with each other, and all results agree well within 1$\sigma$ of each other.
 The worse precision of the frequency-sets of these stars
caused the  worse precision of the results for these stars; the best precision is obtained when the
frequencies are explicitly corrected for the surface term, but as before there 
there could be a larger systematic error in this case, which we have no means of evaluating
independently.

\begin{figure*}
\centerline{
\includegraphics[width=2.25 true in]{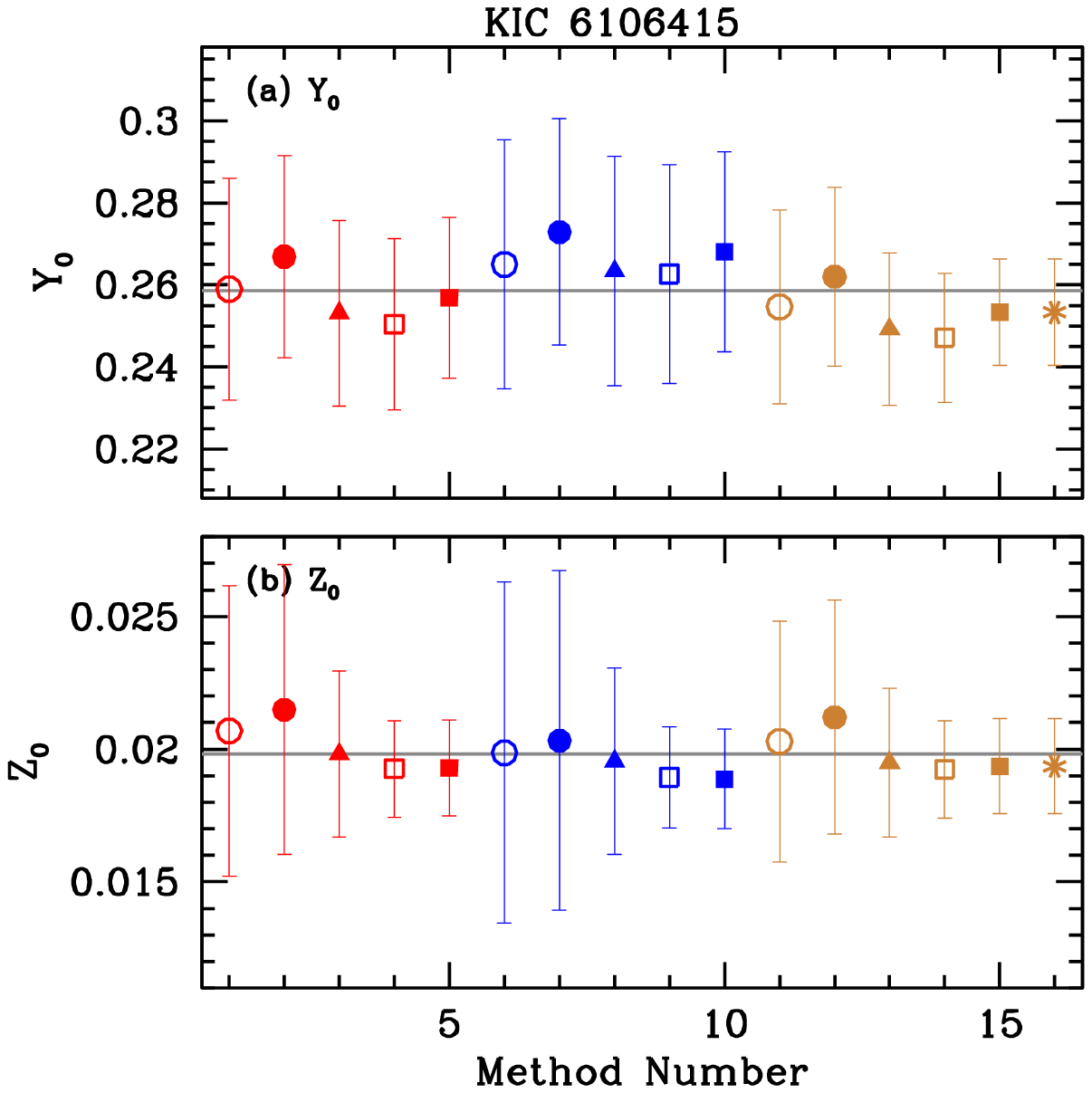}
\includegraphics[width=2.25 true in]{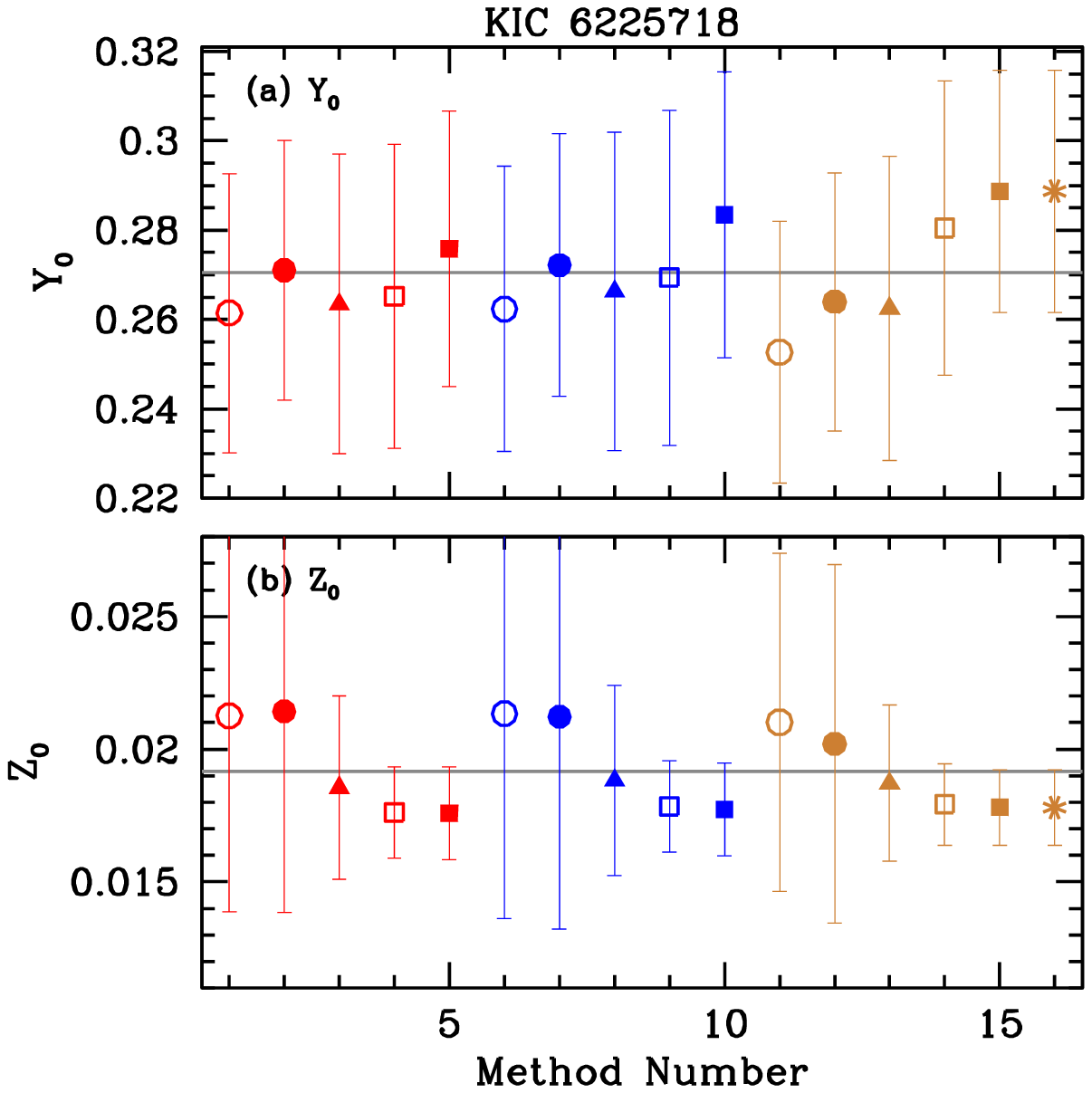}
\includegraphics[width=2.25 true in]{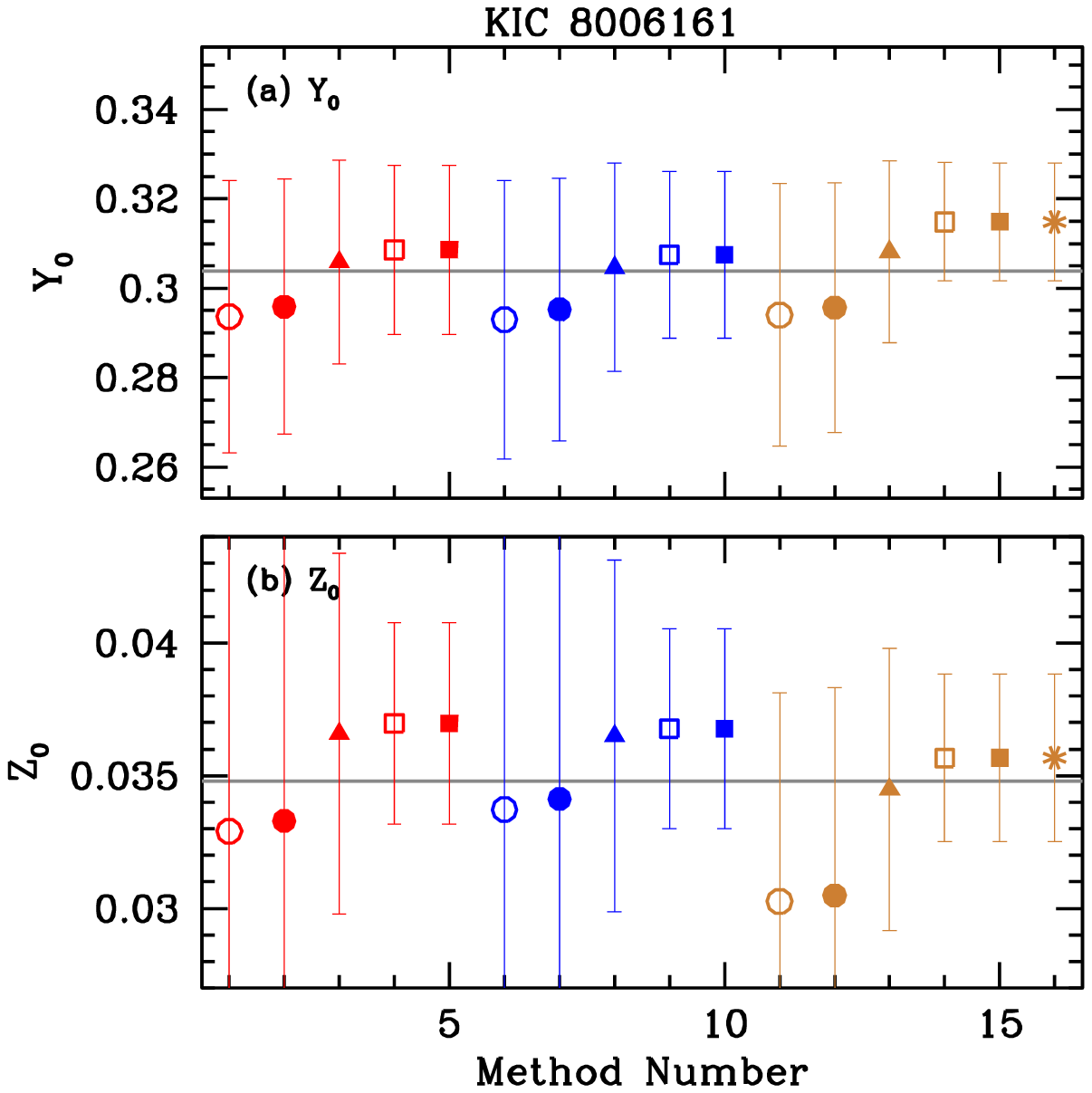}
}
\caption{The initial helium abundance $Y_0$ and initial heavy-element abundance $Z_0$ of
KIC~6106415, KIC~6225718 and KIC~8006161.
The colors are the same as in Fig.~\ref{fig:rma1} The 
horizontal line in each panel is the average of all the values in the panel.
}
\label{fig:sy}
\end{figure*}

Estimates of $Y_0$ and $Z_0$ for KIC~6106415, KIC~6225718 and KIC~8006161 are different though.
The spread in values is much larger (see Fig.~\ref{fig:sy}), particularly for KIC~6225718 and KIC~8006161, the two
stars with comparatively larger frequency uncertainties. However, the results are still the same within 1$\sigma$,
except that $\sigma$ is large. For these stars, meaningful results for $Z_0$ are obtained only for cases where
\teff\ and surface metallicity are explicitly taken into account in the analysis.

\begin{deluxetable*}{lccccc}
\tablecolumns{6}
\tablecaption{Spread in results (in percent) caused by different
ways of handling the surface term. The Method numbers are the same as those in 
Table~\ref{tab:spread}}
\tablehead{\colhead{Methods}& \multicolumn{5}{c}{\% Spread in stellar-parameter estimates} \\
\colhead{compared} & \colhead{Mass} & \colhead{Radius} & \colhead{Age} & \colhead{$Y_0$} & \colhead{$Z_0$}
\label{tab:spread}}
\startdata
1, \phantom{1}6, 11 & 1.1& 0.45 & 2.7 & 2.6 & 6.1\\
2, \phantom{1}7, 12 & 1.1& 0.45 & 2.8 & 2.5 & 6.3\\
3, \phantom{1}8, 13 & 1.0& 0.40 & 2.9 & 2.9 & 2.9\\
4, \phantom{1}9, 14 & 1.2& 0.46 & 1.9 & 2.5 & 1.4\\
5, 10, 15 &           1.1& 0.42 & 1.9 & 2.3 & 1.3\\
\enddata
\end{deluxetable*}

The spread in the global parameters obtained using the different ways of accounting for the
surface term is listed in Table~\ref{tab:spread}. The different rows correspond to
analyses when different non-seismic constraints were used. The first row, for example,
looks at the spread in results when no non-seismic constraint was used.
As can be seen, the change in radius
is typically less than 0.5\% and the change in mass is about 1\%. These numbers
are smaller than the typical uncertainties in mass and radius caused by uncertainties
in asteroseismic and spectroscopic data {  \citep[see e.g.,][]{metcalfe2014,
victor2015, victor2017}} The effect on ages is 
larger and depends on whether or not \teff\ and \feh\ have been use, and the effect is
at the 2-3\% level, much smaller than the usual uncertainties in age.
The effect on $Y_0$ is very similar to that on age. As far as $Z_0$ is concerned, the 
effect of the surface term appears to be tied to whether or not \feh\ is used as a constraint,
and the spread can be as high as 6\% when \feh\ is not used to constrain the results.
We should emphasize that the seeming insensitivity of the results to the way the
surface term is accounted for 
does not mean that we have the option to ignore it altogether. 
{ When a sufficient number of modes is available (as is the case of the 
six stars that have been showcased in this investigation) the $\chi^2$ for the
best fit to the frequencies without a surface term correction is very large. 
This occurs because models that give a good fit to the low-frequency
modes, do not fit the high frequency modes, and vice versa. Models that give better
fits to the low-frequency have properties that are closer to that
of the star than models that fit the high frequency end. This is
a result of the fact that the change in frequency due to the surface term
increases with mode frequency. 
The more problematic case is when modes are available only over a small frequency range (2--3 \dnu)
around \numax, in that case one can get a good fit in terms of $\chi^2$, but 
the chances of fitting a model with wrong parameters is high.
Thus without a proper accounting of the surface term, the results obtained will not be reliable.}

In this work we have not tested the effect of systematic errors in \teff\ and \feh. 
However, a simple analyses of the solar case shows that in the event that the systematic 
error in temperature is of the order of uncertainties in temperature, the effect on mass is no more than
1\%, and the effect on radius is much lower. The effect in age however, can be larger, of the
order of a few per cent. The effect of metallicity errors is slightly larger.
These effects are being examined in detail
in a separate investigation (Bellinger et al. {\it in  preparation}).

\section{Discussion and Conclusions}
\label{sec:conclu}

We have demonstrated that how the surface term in asteroseismic data is accounted for does not affect
estimates of the mass, radius or age of a star. No matter how the surface term is dealt with,
whether by bypassing the term issue completely by using \eps, minimizing its contribution
using ratios, or explicitly removing it using a model for the surface term,  we get
very consistent results for the most important global properties of the stars.
The spread in mass because of the different ways of constraining the surface term
is only about 1\%, the spread is radius is less than 0.5\%, while the spread in age, when \teff\ and \feh\
is used is about 2\%. All these are lower than the typical uncertainties in this 
quantities caused by data errors.
Meaningful estimates of initial helium and metallicity 
require a knowledge of the effective temperature and current metallicity of the stars;
once the spectroscopic data are taken into account, the results are robust with small
enough uncertainties to make the results meaningful.

{  We used all available radial, dipole and quadrupole modes for the stars studied in this
paper.  We did not use
octopole modes even when available since they can observed in very few stars from photometric data.}
 The robustness of the results to surface term corrections hold as long as we have enough
modes of each degree to determine if \eps\ has a degree-dependence, which means that we
need modes of at least two different degrees, or that we have enough modes to calculate
a sufficient number of frequency ratios $r_{01}$ and $r_{02}$. { Mode sets with as few as four to five 
orders per degree give good results.
The range of frequencies covered by the available modes
available does not matter for the \eps-matching or the ratios method, however, it is important
when the surface term is calculated explicitly. The  \citet{ball2014} formulation and 
the  \citet{hans} formulation can only be fitted over a limited frequency range. }
 { The very low frequency range where surface term
corrections are small ($ \lesssim 2000\; \mu$Hz for the solar case, see Fig.~\ref{fig:ssmsurf}),
or at high frequencies where the surface term turns over ($\gtrsim 4000\; \mu$Hz for the
solar case), are fitted badly by these forms.}
For these regions other forms { of the surface term  need to be used. One can, for instance,
explicitly use the scaled solar surface term as was done by the ``ASFIT'' and ``YMCM'' analyses
described in Appendix~A of \citet{victor2015}.}
 However, with photometry measurements, the likelihood of
obtaining { frequencies so low that the functional forms do not apply is small since the
signal from granulation masks the modes, and high-frequency modes usually have large enough
uncertainties that they do not play much of a role in determining stellar properties.}

 All six stars in our sample are main sequence stars. A valid question to ask is what happens to 
more evolved stars, particularly to subgiants. { We do not expect the results to be significantly different
from what we have found for the main sequence stars. 
The p-mode frequencies of subgiants are affected by the surface term the same way as 
the p modes of main sequence stars.
 Given that the radius and mass of stars can be estimated precisely with only average
asteroseismic parameters \dnu\ and \numax \citep[see e.g.][]{ning}, the 
pure p modes of subgiants stars are enough to determine radius and mass precisely,  
and thus these quantities will should only be affected to the same
extent as those for main sequence stars.}
Age is a somewhat different issue. If we look for
age precision at the same level as those for main sequence stars, then the results 
of this work should apply even for evolved stars. However, much more precise ages can be 
obtained for subgiants if mixed-mode frequencies are used \citep{metcalfe2010, deheuvels}.
The frequency of mixed modes is a very sensitive function of age,
and thus the broad-brush statistical technique is not the best if we are to determine the
age of a subgiant to the precision that is possible. To get the better results, we need to start
with the best-fit stellar parameters that fit the p-modes and search around that. Since mixed modes 
have high inertia, and thus less affected by surface effects, we expect the final result to
be insensitive to how the surface term in p modes was removed. 
However, on the red giant branch where
all dipole modes are mixed modes, the usual surface term corrections may not apply at all
\citep{ball2018} and the situation will be very different, unless we confine ourselves to only
using the radial and quadrupole modes. { The expected results for subgiants and
red giants need to be tested properly with real data.}

It should be noted that our results pertain to the effects of the surface term alone, the
model dependence of the different estimates is beyond the scope of this paper. In particular,
one needs to keep in mind that age estimates of stars are always model dependent.

\acknowledgments We would like to than the referee for constructive comments that have
helped in improving this paper. SB would like to acknowledge partial support from  NSF grant AST-1514676 and NASA grant
NNX16AI09G.

\facility{{\it Kepler}}
\software{ YREC \citep{demarque2008}}


\end{document}